\documentclass[twocolumn,prb,aps,amsmath,amssymb,floatfix,showpacs]{revtex4-1}

\usepackage{color}
\usepackage{graphicx}
\usepackage{dcolumn}
\usepackage{bm}
\usepackage{subfigure}

\begin{document}
\newcommand{\bo}{\boldsymbol}
\newcommand{\boq}{\mathbf{q}}
\newcommand{\bok}{\mathbf{k}} 
\newcommand{\bor}{\mathbf{r}}
\newcommand{\boG}{\mathbf{G}}
\newcommand{\boR}{\mathbf{R}}

\title{Density functional perturbation theory for gated two-dimensional heterostructures: Theoretical developments and application to flexural phonons in graphene}

\author{Thibault Sohier$^{1}$}
\author{Matteo Calandra$^{2}$}
\author{Francesco Mauri$^{3,4}$}

\affiliation{
$^{1}$Theory and Simulation of Materials (THEOS), and National Centre for Computational Design and Discovery of Novel Materials (MARVEL), \'Ecole Polytechnique F\'ed\'erale de Lausanne, CH-1015 Lausanne, Switzerland \\
$^{2}$IMPMC, CNRS, Universit\'e P. et M. Curie, 4 Place Jussieu, 
75005 Paris, France\\
$^{3}$Departimento di Fisica, Universit\`a di Roma La Sapienza, 
Piazzale Aldo Moro 5, I-00185 Roma, Italy \\
$^{4}$Graphene Labs, Fondazione Istituto Italiano di Tecnologia
}

\date{\today}

\begin{abstract}
The ability to perform first-principles calculations of electronic and 
vibrational properties of two-dimensional heterostructures in a field-effect 
setup is crucial for the understanding and design of next-generation devices. 
We present here an implementation of density functional
perturbation theories tailored for the case of two-dimensional
heterostructures in field-effect configuration. Key ingredients are
the inclusion of a truncated Coulomb interaction in the direction
perpendicular to the slab and the possibility of simulating charging
of the slab via field-effects. With this implementation we can access
total energies, force and stress tensors, the vibrational properties
and the electron-phonon interaction.
We demonstrate the relevance of the method by studying flexural 
acoustic phonons and their coupling to electrons in graphene doped by field-effect.
In particular, we show that while the electron-phonon coupling to
those phonons can be significant in neutral graphene, it is
strongly screened and negligible in doped graphene, in disagreement 
with other recent first-principles reports. Consequently, the gate-induced 
coupling with flexural acoustic modes would not be detectable 
in transport measurements on doped graphene. 
\end{abstract}

\maketitle


\section{Introduction}
Density Functional
Theory\cite{Hohenberg1964a,Kohn1965,Ihm1979,Martin2004,Engel2011,Parr2015} (DFT)
based on plane-wave basis sets, Kohn-Sham equations and 
pseudopotentials, has proven to be a valuable 
tool to understand and predict electronic and structural properties of materials. 
DFT  provides support in the process of understanding 
and controlling experimentally observed phenomena\cite{Ferrari2006a}.
Inversely, it can be used\cite{Mounet} to identify interesting
new two-dimensional (2D) compounds, thus encouraging their experimental study.
The wide range of potential applications and fascinating phenomena
offered by 2D materials would benefit from accurate DFT simulation
in the 2D framework.

In general, the response of a material 
to a long wavelength periodic perturbation is highly dependent on dimensionality. 
We have recently shown the importance of working in the appropriate 2D framework 
for the computation of dielectric responses\cite{Sohier2015}, optical phonons 
dispersions\cite{Sohier2017a}, and electron-phonon coupling in polar 
materials\cite{Sohier2016}. 
Furthermore, a particularly relevant aspect of 2D materials is their sensitivity 
to external perturbations like external electric fields, as illustrated
by the large gate-induced doping achievable for a 2D material in a 
field-effect transistor (FET) setup\cite{Novoselov2004,Bolotin2008,Efetov2010}.
While the field effect is not part of the 2D material per se, 
its omnipresence in experimental setups and devices motivates the need
to simulate it.

However, current implementations of DFT with three-dimensional 
periodic boundary conditions (3D PBC) are not adequate to the simulation of 
2D materials doped in the FET setup. 
This is mainly due to two reasons. 
The first one pertains to 2D systems in general. 
In the response to long wavelength perturbations 
there is a spurious interaction 
between the system and the out-of-plane
periodic images due to the $1/q^2$ behavior of the Fourier transform
of the 3D Coulomb interaction. This effect becomes relevant for perturbations 
of in-plane momenta comparable or smaller than $\approx 2\pi/c$, where $c$ is the
distance separating the periodic images. This range of momenta is relevant for
electric transport. Additional long range interactions can arise if the slab 
presents a finite dipole along the z-axis. 

The second reason relates to the treatment of charged systems. 
Charging of 2D materials is usually
simulated via the use of a compensating jellium background. This approach is
inappropriate as it represents a uniform doping of a 2D flake and it
reproduces neither the strong voltage drop in proximity of its 
surface\cite{Brumme2014,Brumme2015} nor the asymmetric nature of the FET 
configuration. This asymmetry is precisely the feature that is 
challenging to simulate because it breaks 3D PBC. 
Currently these aspects have not yet been taken into
account in the calculation of vibrational properties via density
functional perturbation theory.

Some methods have been proposed to deal with the FET setup at the DFT
level for the calculation of total energies and forces.
A dipole correction \cite{Neugebauer1992,Bengtsson1999,Meyer2001a,Brumme2014} 
can be used to recover 3D PBC in systems with an out-of-plane dipole moment.
This method has been used to simulate chloronitrides \cite{Brumme2014}
and transition-metal dichalcogenides \cite{Brumme2015} in a FET setup.
Another approach to solve these issues is 
the effective screening medium (ESM) technique \cite{Otani2006}.
The Poisson equation is solved without 3D PBC, resulting in the correct potentials.
The potentials are then modified where the electron density is negligibly small
to allow their use in the KS equations with 3D PBC.
This method has recently been used to simulate a graphene-based
vertical field-effect tunneling transistor \cite{Wang2015} at the DFT
level. However, linear response theory has not been developed 
for any of these methods.
Here we solve this problem and develop a density functional
perturbation theory approach tailored for 2D materials and
heterostructures that includes the possibility of simulating
vibrational properties in FET configuration. 

We use our developments to study flexural phonons in field-effect doped graphene. 
In isolated graphene, those phonons do not couple to electrons 
due the mirror symmetry with respect to the graphene plane\cite{Manes2007}. 
In a FET setup, however, this symmetry is broken and flexural phonons have 
recently been suggested\cite{Gunst2017} as a significant scattering mechanism in 
transport measurements. We recover the expected result
that a symmetry-breaking electric field can activate a 
significant {\it bare} electron-phonon interaction 
with flexural phonons. However, we show that this interaction is 
strongly screened by the electrons and becomes negligible with respect 
to the coupling with in-plane phonons in doped graphene.

This paper is structured as follows. We first describe a model 
for 2D materials doped in the FET setup in Sec. \ref{sec:2DFET}. 
We restrict ourselves to a description in terms of potentials. 
In Sec. \ref{sec:perio_im}, we highlight the issues raised by the presence 
of periodic images to simulate 2D materials in the FET setup, 
and show how the 2D Coulomb cutoff technique can solve those issues.
In Sec. \ref{sec:implementation}, we detail the implementation of the 
2D Coulomb cutoff for the potentials, total energy, forces, phonons and electron-phonon interactions in the Quantum ESPRESSO (QE) distribution\cite{Giannozzi2009}.
Finally, we exploit our implementation of the 2D Coulomb cutoff to study some 
properties of 2D materials specific to the FET setup. Namely, we focus on 
out-of-plane acoustic (ZA) phonons in a graphene FET. We show the 
emergence of a finite phonon frequency at $\mathbf{\Gamma}$ as well 
as a finite coupling to electrons for the ZA phonons, 
two quantities that are zero by symmetry for isolated graphene 
without electric field.

\section{Description of a 2D material doped in the FET setup}
\label{sec:2DFET}

In this section we present our model for a 2D material doped in the FET setup, 
focusing on the potential of such a system. 
The central object is the 2D material itself.
We consider a system with periodicity in the $\{x, y\}$ plane, defined as the infinite
periodic repetition in the plane of a unit cell. 
The positions of the cells are $\mathbf{R}_p= m_1 \mathbf{b}_1+ m_2 \mathbf{b}_2$, where
$m_1$ and $m_2$ are two integers. The primitive lattice vectors $\mathbf{b}_1,\mathbf{b}_2$ have 
coordinates in the $\{x, y\}$ plane.
The $z$-component of $\mathbf{R}_p$ is a constant.
The position of atom $a$ within the unit cell is labeled $\mathbf{d}_a$. 
The atomic internal coordinates $\mathbf{d}_a$ can have different $z$-components such
that all atoms are not necessarily on the same plane, e.g. in the case of multilayered 2D materials.
In reciprocal space, the crystal is described by reciprocal vectors $\mathbf{G}_p$, generated 
by two in-plane primitive reciprocal lattice vectors $\mathbf{b}^*_1$ and $\mathbf{b}^*_2$.

Within the DFT framework, 
the ground state properties of the system are determined by the ground-state 
electronic density $ n(\mathbf{r}_p, z) $, where we separate in-plane (
$\mathbf{r}_p$) and out-of-plane ($z$) space variables, 
as they clearly have different roles in a 2D system:
\begin{align}
n(\mathbf{r}_p, z) &= 2 \sum_{\mathbf{k},s} f(\varepsilon_{\mathbf{k},s}) |\psi_{\mathbf{k},s}(\mathbf{r}_p,z)|^2  \\
\psi_{\mathbf{k},s}(\mathbf{r}_p, z) &= 
\mathbf{w}_{\mathbf{k},s}(\mathbf{r}_p, z) e^{i \mathbf{k} \cdot  \mathbf{r}_p}.
\end{align}
The in-plane wave vector $\mathbf{k}$ and the band index $s$  
define an electronic state.
The Bloch wave functions $\psi_{\mathbf{k},s}$ are the solutions of the 
Kohn-Sham\cite{Kohn1965} (KS) equations.
The KS potential of the 2D system is the sum of the external potential $V^{\rm{2D}}_{\rm{ext}}$ 
(which, for now, consists of the potential generated by the ions $V^{\rm{2D}}_{\rm{ion}}$),
the Hartree potential $V^{\rm{2D}}_{\rm{H}}$, 
and the exchange-correlation potential $ V^{\rm{2D}}_{\rm{XC}}(\mathbf{r}_p, z)$: 
\begin{align}
V^{\rm{2D}}_{\rm{KS}}(\mathbf{r}_p, z)=V^{\rm{2D}}_{\rm{ext}}(\mathbf{r}_p, z) + V^{\rm{2D}}_{\rm{H}}(\mathbf{r}_p, z)
+  V^{\rm{2D}}_{\rm{XC}}(\mathbf{r}_p, z).
\label{eq:KS_pot}
\end{align}
The above quantities possess the 2D-periodicity of the crystal.
That is, for any $f$ 2D lattice-periodic function such as
$n, V^{\rm{2D}}_{\rm{KS}}, V^{\rm{2D}}_{\rm{ext}}, V^{\rm{2D}}_{\rm{H}}$ or $V^{\rm{2D}}_{\rm{XC}}$, 
we can write
\begin{align} \label{eq:2DPBC}
f(\mathbf{r}_p+\mathbf{R}_p, z) = f(\mathbf{r}_p, z) \ .
\end{align}
The 2D Fourier transform of a 2D lattice periodic function reads
\begin{align}\label{eq:2DFT}
f(\mathbf{G}_p, z)&= \frac{1}{S}\int_S  
f(\mathbf{r}_p, z) e^{-i \mathbf{G}_p \cdot \mathbf{r}_p} d \mathbf{r}_p , 
\end{align}
where the integral is over the area of the unit cell $S$. In-plane averages 
are defined as $f(\mathbf{G}_p=0, z)=\langle f \rangle_p (z)$.
In plane averages also extend in the out-of-plane direction. 
A relevant length scale for the out-of-plane extension
of the 2D material would be the electronic density's thickness t, defined such that:
\begin{align}
\int^{t/2}_{-t/2} \langle n \rangle_p(z) \ dz & \approx  n_{0} \ ,
\end{align}
where $n_0 \times S$ is the number of valence electrons per unit cell 
in the system, equal to the sum of the ionic charges $\sum_a Z_a$ in the neutral case. 
The total energy, forces, phonons and electron-phonon interactions of a neutral 2D 
material can be computed using the usual DFT 
formalism\cite{Hohenberg1964a,Kohn1965,Ihm1979,Martin2004,Engel2011,Parr2015} 
based on space integrals of products 
between the electronic density and various potentials. It is then sufficient to carry the out-of-plane integrals over a slab of thickness greater than $t$.

\begin{figure}[h]
\centering
\includegraphics[width=0.47\textwidth]{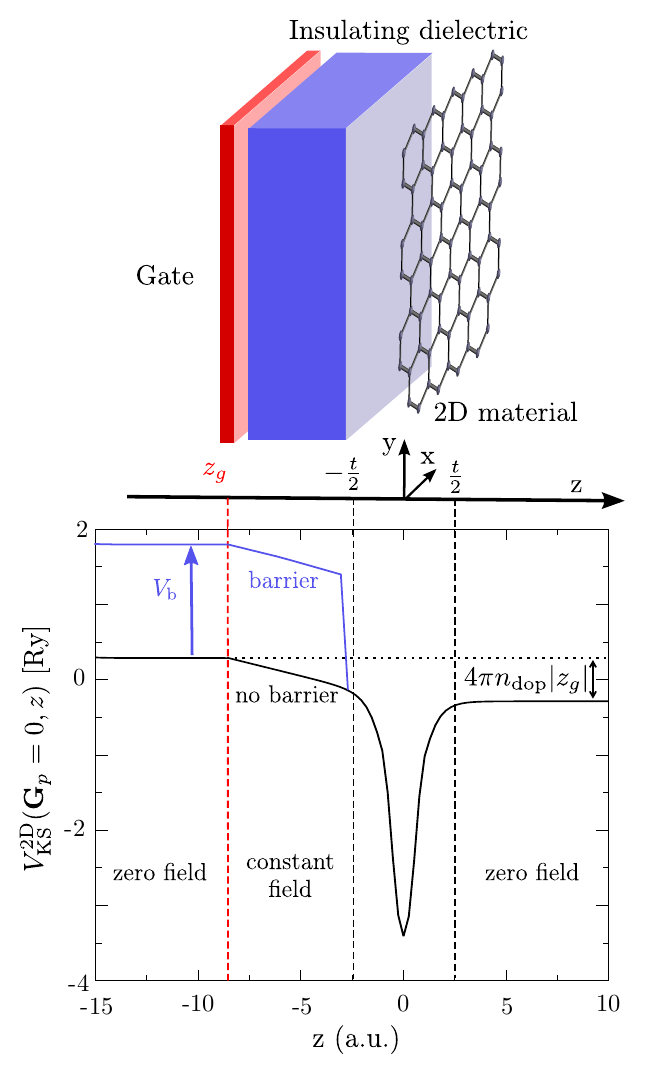}
\caption{The sketch shows a minimal model of the FET setup containing the 2D material and a gate separated by an insulating dielectric.
The plot shows the out-of-plane behavior of the corresponding
Kohn-Sham (KS) potential for a hole-doped single-layer 
2D material (graphene) in the FET setup.
A charged plane simulates the gate. On the left of the material, 
we represent in black the KS potential when only vacuum separates 
the gate and the material. In blue, we add a potential barrier to 
simulate the dielectric material.}
\label{fig:FET-draw}
\end{figure}
We now consider what we must do to simulate this 2D material doped in conditions
emulating the FET setup. We consider a single-gate configuration, as shown 
in Fig. \ref{fig:FET-draw}.
A planar gate is placed parallel to 2D material and a voltage difference 
is applied between the two. 
An insulating material (gate dielectric) separates the 2D material and the gate, 
such that no current can 
flow between them and opposite surface charges accumulate on both sides. 
The key feature of the FET setup is its asymmetry in terms of electric field. 
Between the gate and the 2D material, the electric field is finite. 
On the other side of the 2D material, the electric field is 
zero. In the out-of-plane direction, it is essential that we simulate the correct 
2D potentials in a region at least as large as the thickness $t$.
We will not model every ions and electrons outside this region, 
in the gate-dielectric, substrate or gate. 
We rather propose ways to simulate the effects of
those components on the 2D material.

The main purpose of the FET setup is to charge the 2D material. 
We consider an electron density such that:
\begin{align}
\int  \langle n \rangle_p(z) \ dz = n_0= \sum_a \frac{Z_a}{S} +n_{\rm{dop}} \ ,
\end{align}
where $Z_a$ is the number of pseudo charges of atom $a$, $S$ is the
surface of the 2D unit cell and $n_{\rm{dop}} \times S $ is 
the number of electrons added per unit cell. 
The total charge density of the 2D material is obtained by integrating the sum of 
the charge distributions associated with the ions and electrons:
\begin{align}
\int \langle \varrho_{\rm{ion}}+\varrho_{\rm{elec}}\rangle_p(z) 
dz &= -e  n_{\rm{dop}}.
\label{eq:doped2Dsys}
\end{align}
where the charge densities $\varrho$ are related to the corresponding 
electrons or ions density as $\varrho_{\rm{ion}}=en_{\rm{ion}}$ and $\varrho_{\rm{elec}}=-en$.
In the FET setup of Fig. \ref{fig:FET-draw}, 
the doping comes from the presence of accumulated
counter-charges in the gate. We thus 
add a charged plane of opposite surface
charge density $+e n_{\rm{dop}}$ at $z_{\rm{g}}<-t/2$, 
playing the role of the gate:
\begin{align} \label{eq:rhogate}
\varrho_{\rm{gate}}(z) &= + e  n_{\rm{dop}} \delta(z-z_{\rm{g}}) \\
V^{\rm{2D}}_{\rm{gate}} (z) &= 
+2 \pi e^2 n_{\rm{dop}}  |z-z_{\rm{g}}| . \label{eq:Vgate}
\end{align}
We now have a globally neutral system:
\begin{align}
 \int \langle \varrho_{\rm{ion}}+\varrho_{\rm{elec}}\rangle_p(z) 
+ \varrho_{\rm{gate}}(z)  \ dz = 0 .
\end{align}
The potential of the gate is included in the external potential:
\begin{align}
V^{\rm{2D}}_{\rm{ext}}( \bor_p, z)= V^{\rm{2D}}_{\rm{gate}} (z) +  \
V^{\rm{2D}}_{\rm{ion}}  ( \bor_p, z),
\end{align}
The resulting planar-averaged KS potential $V^{\rm{2D}}_{\rm{KS}}$ 
is the black line in Fig. \ref{fig:FET-draw}, in the case of a hole-doping
monolayer material.
At this point, it has the features expected from a FET setup with 
vacuum as the insulating dielectric. The general characteristics of 
this potential are easily deduced from a simple 
parallel plate capacitor model:
(i) outside the system, the electric field is zero and the potential is constant;
(ii) between the 2D material and the gate, the electric field is constant and the potential is linear with a slope of $4\pi e^2 n_{\rm{dop}}$;
(iii) this electrostatic configuration translates into an out-of-plane dipolar moment which induces a shift in the KS potential:
\begin{align}
\langle V^{\rm{2D}}_{\rm{KS}} \rangle_p (+\infty)-\langle V^{\rm{2D}}_{\rm{KS}} \rangle_p (-\infty)= 4\pi e^2 n_{\rm{dop}} |z_{\rm{g}}| ,
\end{align}
as represented in Fig. \ref{fig:FET-draw}.

The other element to consider to have a minimal working model
for the FET setup is the dielectric separating the gate and the material. 
Its necessity is obvious in the case of electron-doping. 
In that situation, the polarity of the system
pictured in Fig. \ref{fig:FET-draw} is reversed. 
This means that the gate lies at a lower potential
than the 2D material. In our simulations, there would then 
be some leaking of electrons towards the
gate. This is not physical
\footnote{We could think about cold emission (an electron being emitted from a metal 
plate towards an other under a strong electric field), but in that case
we would have to account for the work function of the gate.  
}. 
In a FET setup, this is prevented by the presence of an 
insulator between the gate and the material.
From a more mechanical point of view, the necessity of the dielectric 
is in fact more general. 
Indeed, both for hole- and electron-doping, there is an attractive force between 
the gate and the material, which is simply the electrostatic attraction between two oppositely charged plates:
\begin{align} \label{eq:F_capa}
|\mathbf{F}_{\rm{gate-material}}| 
=   S \times 2\pi e^2 n^2_{\rm{dop}}.
\end{align}
In this context, the dielectric provides a counteracting repulsive force.
To emulate both the insulating and repulsive roles of the dielectric, 
we add a potential barrier in between the material and the gate:
\begin{align} \label{eq:Vbarrier}
V^{\rm{2D}}_{\rm{barrier}} (z) &= 
\begin{cases}
V_{\rm{b}} & \text{if    } z < z_{\rm{b}} \\
0                 & \text{otherwise} 
\end{cases}
\end{align}
where $z_{\rm{g}}<z_{\rm{b}}<0$. This potential can be included in the external 
potential $V^{\rm{2D}}_{\rm{ext}}$.
Adding such a barrier results in the 
potential represented by a blue line in Fig. \ref{fig:FET-draw}.
This barrier potential essentially forbids (or makes highly unlikely) 
the presence of electrons for $z < z_{\rm{b}}$,
thus preventing electrons from leaking towards the gate.
Since the electrons cannot go past the barrier, 
and since the ions are strongly attracted by the electrons, 
the barrier repulses the 2D material as a whole.
As will be detailed later, the equilibrium position of the material 
with respect to the barrier can be determined by relaxation of the forces 
in the system.

In the following section, we explain 
how we deal with the periodic images to obtain the 
KS potential we just described in a plane-wave DFT code wit 3D PBC.
Then, we will detail the modifications implemented to 
compute the total energy, forces, phonons and electron-phonon 
interactions for a 2D material doped in the FET setup. 

\section{Treatment of the periodic images}
\label{sec:perio_im}

{\it Ab initio} calculations based on plane-wave basis sets require periodic boundary 
conditions along the three dimensions (3D PBC). 
In this framework, 
periodic images of the 2D system 
are present in the out-of-plane direction. 
Our goal is for each periodic image to be strictly equivalent to the 2D system
presented in the previous section, 
at least within a certain ``physical region" around the 2D material
(for example, within the boundaries of Fig. \ref{fig:FET-draw}). 
In this section, we detail the issues that arise from the use of 3D PBC 
for the simulation of 
doped systems, 
systems with out-of-plane dipolar moment, 
and systems perturbed at long 
wavelengths. 
We then show how the Coulomb cutoff technique can solve those issues.

\subsection{Inadequacy of 3D PBC}

The 3D-periodic system obtained by adding translated copies of the 2D system
generates potentials that are different from the ones described in the previous section.
This comes from interactions between periodic images, due to the combination of their 
potentials while satisfying PBC. The sum of the KS potential from each periodic image
can be written:
\begin{align}
V_{\rm{KS}}(\mathbf{r}_p, z)&=  \sum_{i} V^{\rm{2D}}_{\rm{KS}}(\mathbf{r}_p, z- i c) ,
\end{align}
where $i$ is an integer, and $c$ is the distance between the periodic images. 
$V^{\rm{2D}}_{\rm{KS}}$ is the potential of the 2D system, 
while $V_{\rm{KS}}$ is the one simulated in DFT with 3D PBC. 
In addition to the 2D PBC of Eq. \eqref{eq:2DPBC} 
that $V^{\rm{2D}}_{\rm{KS}}$ already fulfills, 
$V_{\rm{KS}}$ has to fulfill 
the PBC in the third direction:
\begin{align}
V_{\rm{KS}}(\mathbf{r}_p, z+ic)&= V_{\rm{KS}}(\mathbf{r}_p, z) \ \ , \ \forall \ i .
\end{align}

\begin{figure}[h]
\centering
\includegraphics[width=0.45\textwidth]{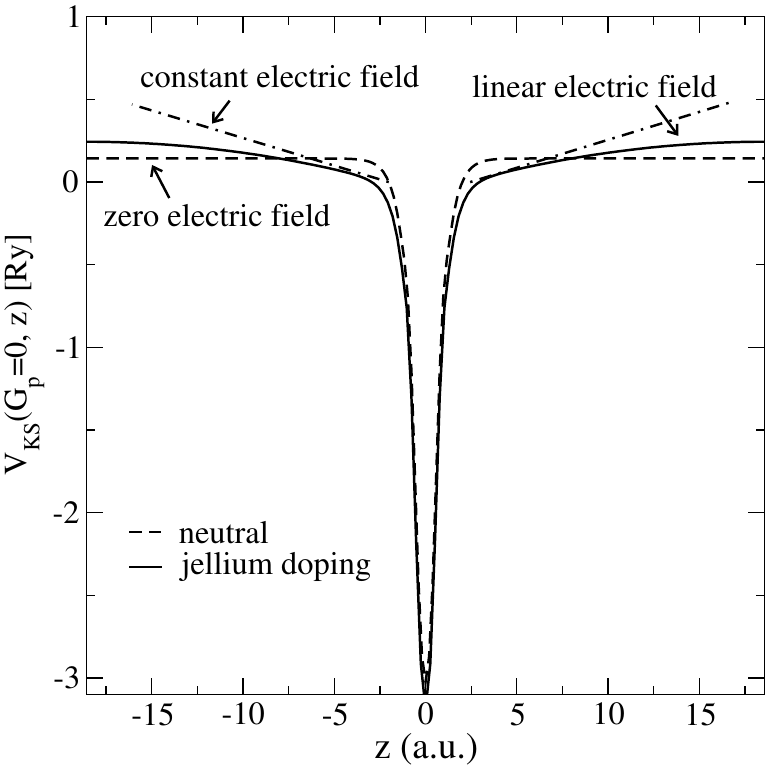}
\caption{ Planar-averaged KS potential in the out-of-plane direction as simulated 
in DFT with 3D PBC for neutral (dashed line) and doped (plain line) graphene. 
In the case of doped graphene, the quadratic behavior of the KS potential indicates 
the presence of a jellium background and a linear electric
field. The dot-dashed line represent the behavior one would expect from an isolated monopole with the same surface charge density as the 2D material 
$V_{\rm{mono}}(z) = - 2 \pi e^2 n_{\rm{dop}} |z|$ .}
\label{fig:jellium_dop}
\end{figure}

We first consider a doped 2D material.
Away from the direct vicinity of the material, this system behaves 
like a monopole, with 
$\lim_{|z| \to \infty} \langle V^{\rm{2D}}_{\rm{KS}} \rangle_p(z) = \infty$,
and $V_{\rm{KS}}$ is obviously ill-defined.
As mentioned before the standard method in current plane-wave DFT 
packages amounts to the use of a jellium background.
Each slab is then globally neutral, containing the doped material and 
a uniform distribution of compensating charges. 
In between the periodic images, the resulting 
potential is quadratic in $z$, with extrema at mid-distance between layers, as shown in Fig. \ref{fig:jellium_dop}. 
This potential does fulfill the PBC 
and doesn't diverge. However, it is quite different from the potential 
one would expect for a charged, isolated 2D material. Indeed, away from the direct 
vicinity of the materials, one would expect to recover a
linear potential, similar to that generated by an isolated monopole. 

Now we consider a 2D system with a global dipolar moment in the out-of-plane 
direction and a $V^{\rm{2D}}_{\rm{KS}}$ potential like in Fig. \ref{fig:FET-draw}.
Here, each 
periodic image is globally neutral. However, the potential 
$\langle V_{\rm{KS}} \rangle_p(z)$ would 
experience a shift with each periodic images, eventually diverging. Imposing PBC forbids 
this kind of situation. Instead, it leads to a combination of additional electric field and 
re-organization of the charge so that the total average electric field in one slab is zero\cite{Neugebauer1992,Bengtsson1999,Meyer2001a}.
Here again, we loose the equivalence with $V^{\rm{2D}}_{\rm{KS}}$. 

Finally, 3D PBC are very problematic when the
system is perturbed at small wave vector. 
If a 2D charge density is modulated according to an in-plane 
wave vector $\mathbf{q}$, it generates a potential decreasing as 
$e^{-|\boq||z|}$ in the out-of-plane direction.
At small wave vector, the extent of the potential induced by the electron 
density is thus very large. When it is of the order of the distance between 
periodic images, there is some spurious interactions.
This issue is critical when simulating the screening properties\cite{Sohier2015} of 
the material as well as its response to phonon 
perturbations\cite{Sohier2016,Sohier2017a}.

\subsection{Isolate the layers with 2D Coulomb cutoff }

To reduce interactions between periodic images, a naive solution is to 
increase the distance between them. 
However, in a plane-waves framework, the cost of the calculations increasing 
linearly with the distance, this method can be
very expensive, especially in the particular cases presented before.
Furthermore, this method inevitably fails for DFPT 
in the long wavelength limit, as spurious interactions are bound to
affect the response of the material at small enough wave vectors. 

One solution for the FET setup and for systems with an out-of-plane dipolar 
moment in general is 
to add a dipole correction to catch up the potential shift 
\cite{Neugebauer1992,Bengtsson1999,Meyer2001a,Brumme2014}.
However, the dipole correction has to be recalculated self-consistently
at each iteration, and this method has not been extended to the DFPT framework.

Here, we tackle 3D PBC issues by using the Coulomb cutoff technique\cite{Jarvis1997,Rozzi2006,Ismail-Beigi2006}, 
successfully used by other codes\cite{Marini2009,Gonze2016} 
in different contexts.
The general idea is to cut all the potentials off between the periodic images. 
In effect, all physical links between periodic images are severed because the potential generated by one periodic image does not reach the others.
Each slab is effectively isolated. There is no {\it physical} 
3D-periodic system anymore. 
There is a 2D-periodic system, copied and repeated in the third dimension
in order to build potentials that {\it mathematically} fulfill 3D PBC.

Each long-range potential ($V\equiv V_{\rm{ion}},V_{\rm{H}},V_{\rm{gate}}$)
in the original 3D code is generated by a certain distribution of charges 
via the Coulomb interaction $v_c(\bor)=\frac{e^2}{|\bor|}$.
To build the corresponding cutoff potentials in the code 
($\bar{V} \equiv \bar{V}_{\rm{ion}},\bar{V}_{\rm{H}},\bar{V}_{\rm{gate}}$), we use the following cutoff Coulomb interaction:
\begin{align}
\bar{v}_c(\bor)=\frac{e^2\theta(l_z-|z|)}{|\bor|} , 
\label{eq:cutoff}
\end{align}
where $\bor \equiv (\bor_p, z) $ is a generic 3-dimensional space variable .
An arbitrary charge density $\varrho$ then generates the following potential:
\begin{align}
\bar{V}(\bor)=  \int \frac{e\varrho(\bor')}{|\bor-\bor'|} \theta(l_z-|z|) d\bor' .
\end{align}
Roughly speaking, considering a single charged plane,
we generate its potential only within a certain slab of thickness
$2l_z$ centered on the charge distribution. Within this slab, we have that 
$\bar{V}(\bor)=V^{\rm{2D}}(\bor)$. Outside of this slab, the potential is zero.
Each periodic image of each 
charge distribution ($ \varrho_{\rm{ion}},\varrho_{\rm{elec}},\varrho_{\rm{gate}}$)
generates its own potential within its own slab.
To fulfill 3D PBC, the simpler way is to cut off midway between the periodic images: 
\begin{align}
l_z=\frac{c}{2}.
\end{align}
Since the potentials $V^{\rm{2D}}_{\rm{ion}}$, $V^{\rm{2D}}_{\rm{H}}$, and $V^{\rm{2D}}_{\rm{gate}}$ are symmetric with respect to the plane of the associated
subsystem (ions, electrons, gate), they have the same value on both sides of the corresponding slab. 
$\bar{V}_{\rm{ion}}$, $\bar{V}_{\rm{H}}$,
and $\bar{V}_{\rm{gate}}$ are each continuous and periodic, and so is their sum
$\bar{V}_{\rm{KS}}$. However, since the slabs of each subsystem do not
coincide, the KS potential is only physical within the overlap of the subsystems' slabs.
\begin{figure}[h]
\centering
\includegraphics[width=0.45\textwidth]{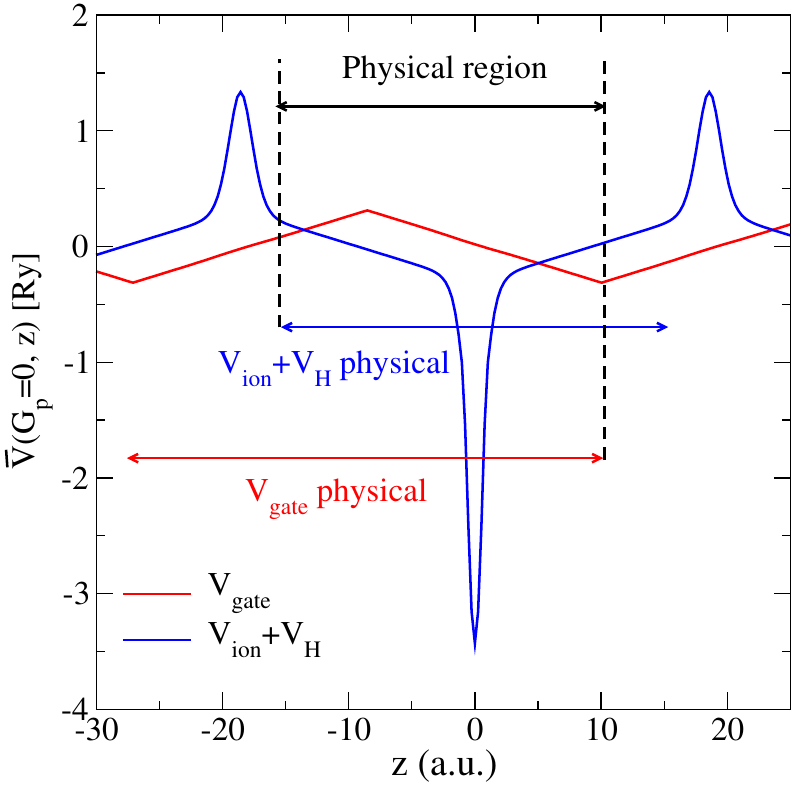}
\includegraphics[width=0.45\textwidth]{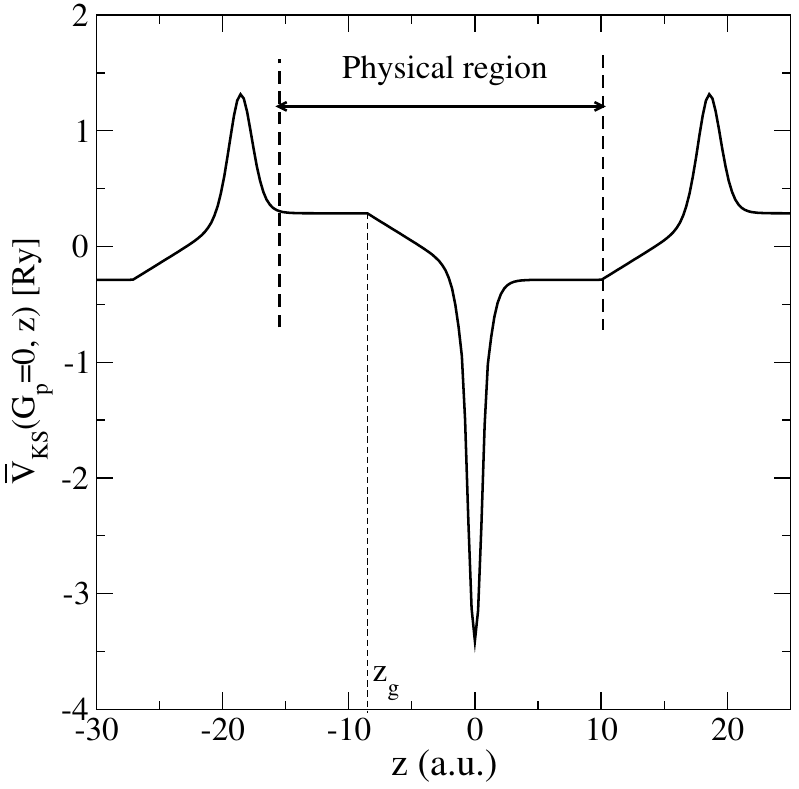}
\caption{Determination of the physical region.
In the upper panel, we show the gate potential and the material's potential 
$\langle \bar{V}_{\rm{ion}}+\bar{V}_{\rm{H}}\rangle_p(z)$, 
and indicate where they make sense physically.
The physical region is the overlap between those regions. 
In the lower panel, we show that the 
KS potential $\langle \bar{V}_{\rm{KS}}\rangle_p(z)$ 
(the sum of the potentials above) 
makes sense within the physical region. 
In both plots, $c\approx 37$ a.u. and $z_{\rm{g}}\approx - 8.5$ a.u.}
\label{fig:Phy_reg}
\end{figure}
This overlap region defines a ``physical region", as illustrated in Fig. \ref{fig:Phy_reg}
, where all the potentials make sense. Outside of this region, there are some
spurious unphysical variations of the KS potential.
Those spurious variations are a necessary consequence of fulfilling 3D PBC.
Let us consider the example Fig. \ref{fig:Phy_reg} in more details.
The simplest subsystem is the gate because $\varrho_{\rm{gate}}$
is infinitely thin in the out-of-plane direction. Within the slab  
$z \in [z_{\rm{g}}-\frac{c}{2} ; z_{\rm{g}}+\frac{c}{2}]$, we see the potential generated by the 
gate at $z_{\rm{g}}$. 
\begin{align}
\bar{V}_{\rm{gate}}(\mathbf{r}_p, z)&=  
\sum_{i} V^{2D}_{\rm{gate}}(\mathbf{r}_p, z- i c) \theta(\frac{c}{2}-|z-i c|) \\
&= V^{2D}_{\rm{gate}}(\mathbf{r}_p, z) \text{  if }  z \in [z_{\rm{g}}-\frac{c}{2} ; z_{\rm{g}}+\frac{c}{2}] .
\end{align}
For $z$ outside of this interval, we see the potential generated by the 
neighboring periodic images of the gate, which has no physical sense with respect to the
2D system represented in Fig. \ref{fig:Phy_reg}.
For the electrons, the charge distribution $\varrho_{\rm{elec}}$ is spread
in the out-of-plane direction. Each infinitesimal slice of electronic density with 
surface charge density $\varrho_{\rm{elec}}(z)dz$ 
generates its contribution to the Hartree potential only within a certain slab.
The Hartree potential is physical only within the overlap of all those slabs.
If the electrons are centered around a position
$z_{\rm{e}}$, that would be
$z \in [z_{\rm{e}}-\frac{c}{2}+\frac{t}{2} ; z_{\rm{e}}+\frac{c}{2}-\frac{t}{2}]$.
The ions are in a similar situation, but the charge distribution is much less spread.
The difference in the spreading of $\varrho_{\rm{elec}}$ and $\varrho_{\rm{ion}}$
leads to the bumps we can observe for $\bar{V}_{\rm{ion}}+\bar{V}_{\rm{H}}$
, at $z \approx \pm 18.5$ a.u. in Fig. \ref{fig:Phy_reg}.
Thus, the unphysical variations of the KS potential outside the physical region 
are due to the addition of potentials generated by incomplete subsystems
or different periodic images. 

Nevertheless, everything happens as in the previous section within the physical region
associated to the KS potential. To simulate the system, we just need to make sure that 
the 2D material lies in this physical region.
We will need 3D Fourier transforms $\bar{V}(\mathbf{G}_p, G_z)$, 
easily related to the 2D Fourier transform of $V^{2D}(\mathbf{G}_p, z)$, Eq. 
\ref{eq:2DFT}, as follows:
\begin{align}
\bar{V}(\mathbf{G}_p, G_z)
&= \frac{1}{c}\int_{-c/2}^{c/2} dz \ 
V^{\rm{2D}}(\mathbf{G}_p, z) e^{-i G_z z} .
\label{eq:2DTF} 
\end{align}

\section{Implementation}
\label{sec:implementation}

In this section we detail the implementation of the Coulomb cutoff for the potentials, 
and show how it affects the energies, forces, phonons and EPC in the code. 
Most of the potentials, or at least their long-range parts, are calculated
in reciprocal space. We thus need the Fourier transform of
Eq. \ref{eq:cutoff}, as defined in Ref. \onlinecite{Rozzi2006}:
\begin{align} \label{eq:CoulCut}
\begin{split}
\bar{v}_c(\mathbf{G}_p, G_z \neq 0) &=
 \frac{4\pi e^2}{ |\mathbf{G}_p|^2+G_z^2}  \times \left[1 - e^{-|\mathbf{G}_p| l_z}  \cos(G_z l_z)  \right], \\
\bar{v}_c(\boG=0) &= 0 .
\end{split}
\end{align}
The choice of the $\boG=0$ value is just a convention since 
every potential is defined
up to a constant. Here, we choose the same convention as in the original 3D code, 
such that the average of a potential over the unit cell is zero.
A more detailed justification about this choice, especially its implications in 
terms of energy, can be found in App. \ref{app:2DQE}. 
For clarity, we will often need to describe the implementation 
of the original 3D code first in order to identify what needs to be modified.
We use different notations to distinguish the potentials that stay as implemented 
in the original 3D code (noted $V$) and those that are modified 
with the implementation of the 2D Coulomb cutoff (noted $\bar{V}$).
For other quantities (energy, forces, phonons and EPC), 
such distinction in the notation is not necessary. 
Indeed, their definition essentially does not change, it is the potential that is used to compute them that changes.

\subsection{KS Potential}

The KS potential is the sum of the external potential, the Hartree potential, and the exchange-correlation potential:
\begin{align}
\bar{V}_{\rm{KS}}(\mathbf{r}_p, z)=\bar{V}_{\rm{ext}}(\mathbf{r}_p, z) + \bar{V}_{\rm{H}}(\mathbf{r}_p, z)
+  V_{\rm{XC}}(\mathbf{r}_p, z).
\end{align}
The exchange-correlation potential is short-range and does not 
need to be cut off. Thus, it will be ignored in the following. 
We note, however, that a great majority of commonly used functionals 
are derived in the framework of the 3D electron gas. 
While their relevance in the context of 2D materials
is obviously questionable, the development of new functionals for 2D materials 
is out of scope for the present work. 
The implementation presented here is valid for all the usual types of
pseudopotentials (norm-conserving, ultra-soft, projector-augmented wave functions).
Indeed, we only modify the long-range parts of the potentials, 
which are independent of the pseudopotential type.

The external potential is the sum of the ionic, gate and barrier potentials:
\begin{align} \label{eq:extPot}
\bar{V}_{\rm{ext}} (\mathbf{r})&= \bar{V}_{\rm{ion}} (\mathbf{r})+\bar{V}_{\rm{gate}} (z)
+\bar{V}_{\rm{barrier}} (z) .
\end{align}

\subsubsection{Ionic Potential}
The ionic potential is separated in local and non-local parts
 $\bar{V}_{\rm{ion}}=\bar{V}^{\rm{loc}}_{\rm{ion}}+V^{\rm{non-loc}}_{\rm{ion}}$.
The non-local part is short-range. It does not need to be cut off and is ignored here.
We need to compute the Fourier transform of the cutoff local potential 
$\bar{V}^{\rm{loc}}_{\rm{ion}}(\boG)$. It is computed from the  
the pseudopotentials, which are separated in short-range and long-range parts. 
We first describe this separation as it is done in the original code, 
identify what must be modified, then present the implementation of the cutoff. 

\underline{In the original 3D code}, the local part of the pseudopotential is a radial function 
in real space $v_a(|\mathbf{r}|)$ associated to each type of atom. 
It is separated in short-range (SR) and long-range (LR) parts:
\begin{align} \label{eq:LR-SR3D}
v_a(|\mathbf{r}|) &= v_a^{ \rm{SR}}(|\mathbf{r}|) + 
v_a^{ \rm{LR}}(|\mathbf{r}|) \\ 
v_a^{ \rm{SR}}(|\mathbf{r}|)&= v_a(|\mathbf{r}|)+ 
\frac{Z_a e^2 \rm{erf}(\sqrt{\eta}|\mathbf{r}|)}{|\mathbf{r}|} \\
v_a^{ \rm{LR}}(|\mathbf{r}|)&= 
- \frac{Z_a e^2 \rm{erf}(\sqrt{\eta}|\mathbf{r}|)}{|\mathbf{r}|}  , 
\end{align}
where $\rm{erf}(\sqrt{\eta} |\mathbf{r}|)$ 
is the error function with $\eta$ as a tuning parameter 
(see App. \ref{app:eta} for more details on that parameter).
The pseudopotential $v_a^{ \rm{SR}}(\mathbf{r})$ is indeed short-range because 
it always behaves as $- \frac{Z_a e^2 \rm{erf}(\sqrt{\eta}|\mathbf{r}|)}{|\mathbf{r}|}$ 
for $|\mathbf{r}|$ large enough. In particular, we have that 
$v_a^{ \rm{SR}}(\mathbf{r}) =0$ for $|\mathbf{r}| \ge r_{\rm{SR}}$.
The Fourier transform of the SR part is calculated via numerical integration, 
while the LR part is analytic.
The SR part, specific to each atom, is Fourier transformed on a finite sphere:
\begin{align}
v_a^{ \rm{SR}}(\mathbf{G})&= 
\frac{1}{\Omega}\int_0^{|\bor|=r_{\rm{SR}}}   
v_a^{ \rm{SR}}(\mathbf{r}) e^{-i\mathbf{G}\cdot \mathbf{r}} \ d\mathbf{r}, 
\end{align}
where $\Omega=S \times c$ is the volume of the unit cell. 
The potential $v_a^{ \rm{SR}}$ does not need to be cut off as long as $r_{\rm{SR}}<l_z$, 
which is easily satisfied.
The Fourier transform of the LR part $v_a^{ \rm{LR}} (\boG)$ is easily 
found analytically, since $v_a^{ \rm{LR}} (|\mathbf{r}|)$ is the potential 
generated by a Gaussian distribution of charges:
\begin{align} \label{eq:pseudoLRG3D}
v_a^{ \rm{LR}}(\mathbf{G})&=
- \frac{Z_a}{\Omega} v_c(\mathbf{G}) e^{-|\mathbf{G}|^2/4\eta} . 
\end{align}
This SR/LR separation is implemented in the original 3D code to 
enable the restriction of numerical Fourier transforms to a 
finite region of space. The original code also relies heavily 
on the rotational invariance of the radial pseudopotentials to define the 
arrays containing their Fourier transforms.

\underline{In our 2D implementation}, we replace the analytic 
LR part of the pseudopotential by its cutoff version:
\begin{align} \label{eq:pseudoLRG}
\bar{v}_a^{ \rm{LR}}(\mathbf{G})&=
- \frac{Z_a}{\Omega} \bar{v}_c(\mathbf{G}) e^{-|\mathbf{G}|^2/4\eta} . 
\end{align}
The SR/LR separation turns out to be very convenient to implement the Coulomb cutoff.
However, since the Coulomb cutoff breaks the rotational 
invariance, it cannot be implemented as a simple 
modification of the existing array. A separate array for 
the cutoff LR part is calculated in a separate routine.
It is then added to the SR part when constructing
the local part of the ionic potential:
\begin{align} \label{eq:Vloc}
\bar{V}^{\rm{loc}}_{\rm{ion}}(\mathbf{G})&= 
 \sum_{a} e^{-i\mathbf{G} \cdot \mathbf{d}_a} \left( v_a^{ \rm{SR}}(\mathbf{G}) + 
\bar{v}_a^{ \rm{LR}}(\mathbf{G}) \right) .
\end{align}

\subsubsection{Hartree Potential}
The Hartree potential is relatively easy to cut off. It is computed in 
reciprocal space from the electronic density:
\begin{align} \label{eq:barVh}
\bar{V}_{\rm{H}}(\mathbf{G}) &= \bar{v}_c(\mathbf{G})  n(\mathbf{G}) .
\end{align} 

\subsubsection{Gate Potential}
The gate potential must be added for a doped system.
In practice, the potential of the gate is added in real space to the external potential.
We define directly in real-space the saw-tooth potential generated by 
$\varrho_{\rm{gate}}$, Eq. \eqref{eq:rhogate} via the cutoff Coulomb 
interaction Eq. \eqref{eq:CoulCut}. Within the interval
$z \in [z_{\rm{g}}-\frac{c}{2},z_{\rm{g}}+\frac{c}{2}]$, it is defined as:
\begin{align}
\bar{V}_{\rm{gate}} (z) &= 2 \pi e^2 n_{\rm{dop}} 
\left( |z-z_{\rm{g}}| - \frac{l_z}{2} \right) , 
\end{align}
where the constant term is due to the definition of $\bar{v}_c(\boG=0)$. 
It sets the out-of-plane average of the potential to zero.
A second gate can be added to provide more flexibility and 
to simulate the combination of bottom and top gate. 
We simply define two separate gate potentials 
(index "bot" for bottom gate and "top" for top gate): 
\begin{align}
\bar{V}^{\rm{bot}}_{\rm{gate}} (z) &= 2 \pi e^2 n_{\rm{bot}} 
\left( |z-z^{\rm{bot}}_{\rm{g}}| - \frac{l_z}{2} \right) \\
\bar{V}^{\rm{top}}_{\rm{gate}} (z) &= 2 \pi e^2 n_{\rm{top}} 
\left( |z-z^{\rm{top}}_{\rm{g}}| - \frac{l_z}{2} \right) , 
\end{align}
and add them together to form the total gate potential:
$\bar{V}_{\rm{gate}} (z)=\bar{V}^{\rm{bot}}_{\rm{gate}} (z)+
\bar{V}^{\rm{top}}_{\rm{gate}} (z) $
where the charges should be such that the whole system is neutral: 
$n_{\rm{top}}+n_{\rm{bot}}=n_{\rm{dop}}$.

\subsubsection{Barrier Potential}
The barrier potential is necessary to relax the forces in the system and
to prevent electrons from leaking towards the gate.
It is also used to prevent electrons from going outside the physical region.
Indeed, the variations of the potential outside the physical region can lead
to the presence of potential wells.
Placing a barrier potential outside the physical region ensures that no unphysical leaking
occurs. 
The barrier is added in real space along with the gate.
In practice, this barrier consists in the addition of a constant to the external potential
within a certain region in the out-of-plane direction. For
$z \in [-\frac{c}{2},+\frac{c}{2}]$, it is defined as:
\begin{align} \label{eq:Vb}
\bar{V}_{\rm{barrier}} (z) &= 
\begin{cases}
V_{\rm{b}} & \text{if    } z < z_{\rm{b}1} \text{  or   } z_{\rm{b}2} < z \\
0                 & \text{otherwise} 
\end{cases}
\end{align} 
The borders of the barrier at $z_{\rm{b}1}$ and $z_{\rm{b}2}$ are smoothed
via a linear transition from $V_{\rm{b}}$ to $0$ on a small distance. 
The implementation of the gate and the barrier was adapted from a previous
modification of the code, discussed in Ref. \onlinecite{Brumme2014}.

\subsubsection{Verifications}
\begin{figure}[h]
\centering
\includegraphics[width=0.45\textwidth]{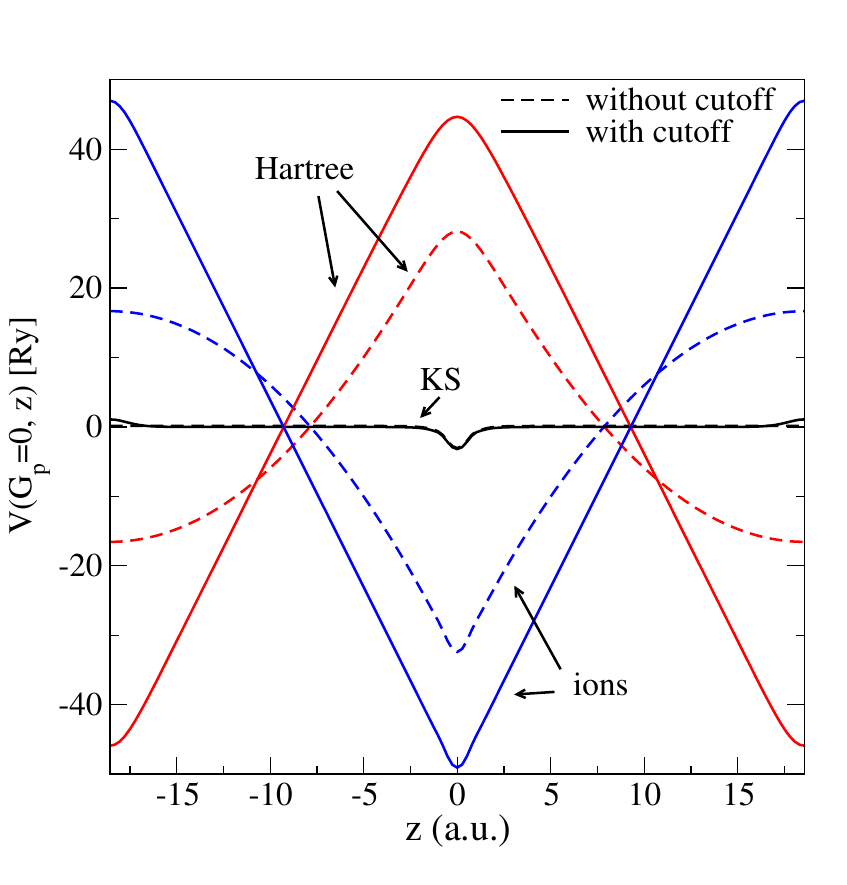}
\includegraphics[width=0.45\textwidth]{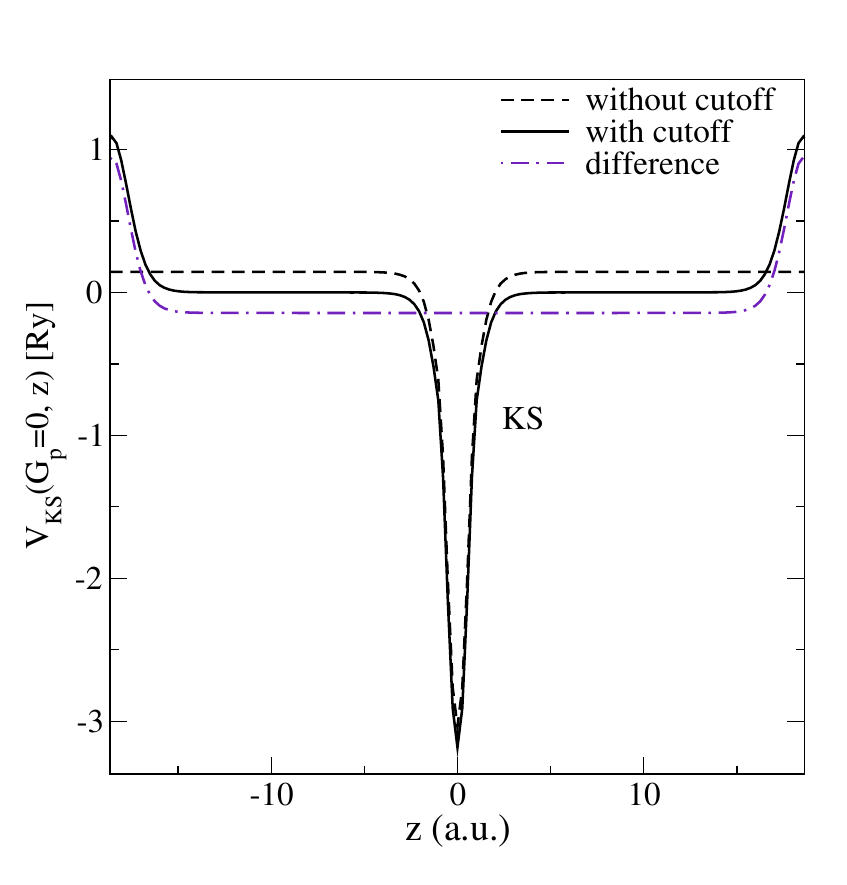}
\caption{In the upper panel, we show the planar-averaged ionic, Hartree and KS potential 
obtained with and without the 2D Coulomb cutoff.
In the lower panel, we zoom in on the KS potential. Within the physical region, 
the KS potentials with and without cutoff coincide up to a constant, 
as is demonstrated by the difference $\langle \bar{V}_{\rm{KS}}-V_{\rm{KS}}\rangle_p (z)$
(dash-dot indigo line).}
\label{fig:potentials}
\end{figure}

To check the consistency of our modifications on the potentials, 
we can first simulate the potentials of 
a neutral and non-polar 2D system, without gate or barrier.
The corresponding ionic, Hartree and KS potentials are plotted with and without
the 2D Coulomb cutoff in Fig. \ref{fig:potentials}.
With 3D PBC, setting the $\boG=0$ value of the ionic or Hartree potential to zero 
is equivalent to the inclusion of a compensating jellium background. 
The potentials we observe then correspond to either ions or electrons
bathed in the associated jellium. 
This leads to a quadratic behavior in $z$ between the 
periodic images. 
When the 2D Coulomb cutoff is applied, we 
recover the linear behavior in $z$. Setting the $\boG=0$ value of the ionic 
or Hartree potential to zero leads to a simple shift.
For such a neutral and non-polar system, the KS potentials with and without cutoff
coincide up to a constant within the physical region. 
This constant comes from the fact that both KS potential average to zero 
but the cutoff KS potential has bumps outside the physical region 
while the other does not. 

\begin{figure}[h]
\centering
\includegraphics[width=0.45\textwidth]{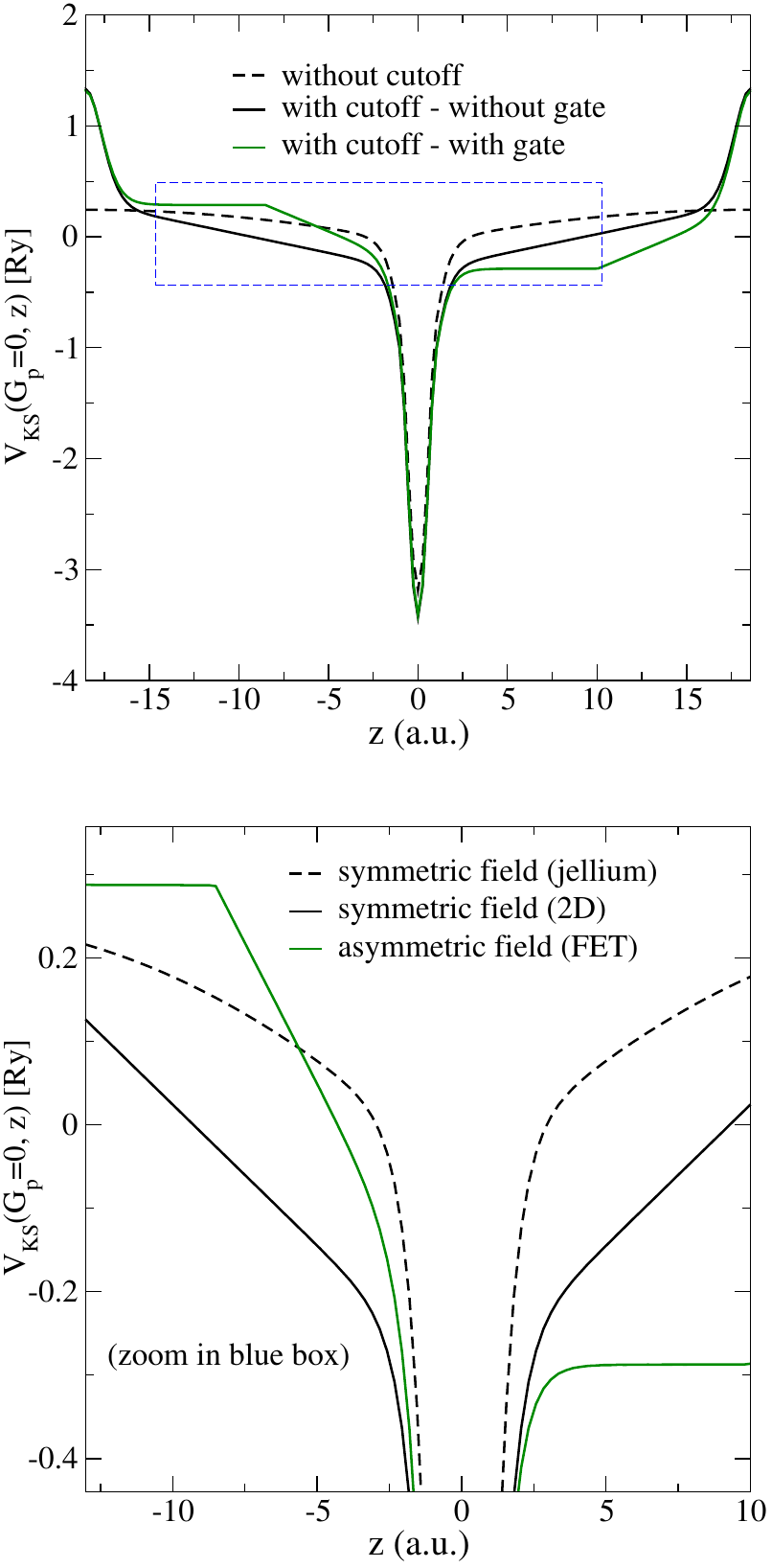}
\caption{The planar-averaged KS potential of hole-doped graphene is simulated in various situations to show the different configurations in terms of electric field.
The label "without cutoff" means that the standard 3D code was used.
The label "with cutoff" means that the 2D Coulomb cutoff was implemented.
In that case, we plotted the result with and without a gate.
The lower panel is a zoom in the region delimited by the 
blue box in the upper panel.
}
\label{fig:ho_dop}
\end{figure}

Let us now simulate the KS potential of a hole-doped 2D material as shown
in Fig. \ref{fig:ho_dop}. Using the original code with 3D PBC, we obtain the potential 
of the material bathed in a jellium compensating for the added charge 
(or missing electrons). In that case, the KS potential is quadratic,
with a varying slope and thus a varying electric field.
The electric field is symmetric with respect to the plane of the 2D material. 
It vanishes midway between the periodic images, 
on the left and right borders of Fig. \ref{fig:ho_dop}. 
If we use the 2D Coulomb cutoff without adding 
a gate, we obtain the potential that would be generated by the doped 
2D material in vacuum, as in Eq. \eqref{eq:doped2Dsys}, within the physical region. 
If we add a compensating charged plane to simulate the gate, 
we obtain the configuration of the FET setup, with a finite electric field on the left of the 2D material, 
and zero electric field on the right.
\begin{figure}[h!]
\centering
\includegraphics[width=0.45\textwidth]{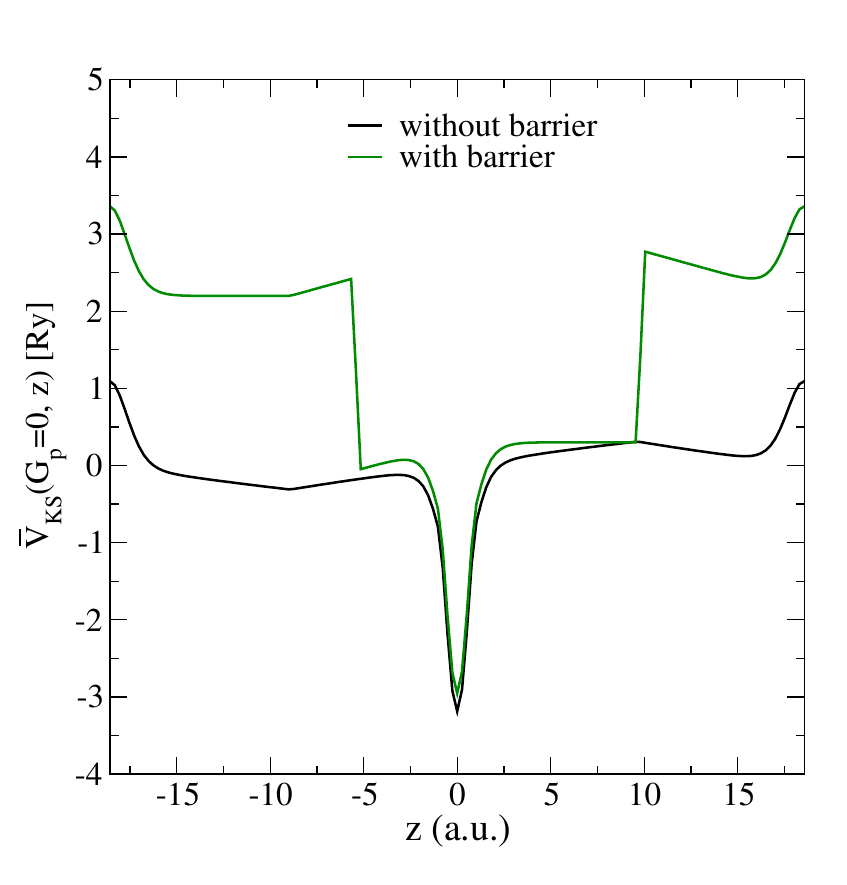}
\caption{
Planar-averaged KS potential in the case of electron doping.
The gate is always included. We plot the potential without barrier to show how there is some 
unphysical leakage. With the barrier, we effectively prevent electrons from reaching any
of the unphysical potential wells.}
\label{fig:el_dop}
\end{figure}

Finally, we simulate the KS potential of an electron-doped system to show the necessity of the 
barrier potential in Fig. \ref{fig:el_dop}. 
Without the barriers, some potential wells appear on both sides of the 2D
material. On the left, this is due to the presence of the positively charged gate. 
On the right, this is due to the unphysical variations of the KS potential outside the physical
region. Electrons leak towards those potential wells, 
which can be inferred here from the slopes of the KS potential in the vicinity of the 
2D material. Compared to what we should obtain in the FET setup, 
the slope of the KS potential on the left of the 2D material is too small while the slope 
on the right is not zero. This is due to the Hartree potential contribution 
from the electrons that leaked in the potential wells. 
This is not what we want to simulate. 
The addition of a potential barrier prevents the electrons
from leaking towards the barrier or outside the physical region, and we find the 
right slopes (or electric field) in the vicinity of the material.

\subsection{Total Energy}

The total energy per unit cell associated with the system is:
\begin{align}\label{eq:TotalEnergy}
E_{\rm{tot}} &= E_{\rm{kin}}+ E_{\rm{ext}}+E_{\rm{H}}+E_{\rm{XC}}+E_{\rm{i-i}} 
 +E_{\rm{g-i}}+E_{\rm{g-g}} .
\end{align}
It is the sum of the kinetic energy of the electrons, 
the energy of the electrons in the external potential, the Hartree energy, the 
exchange-correlation energy, the ion-ion interaction energy, 
the energy of the ions in the potential of the gate(s), and finally, 
the self-interaction energy of the gate(s).
The terms $E_{\rm{kin}}$ and $E_{\rm{XC}}$ are short-range. They
are computed as in the standard 3D code.
The computation of the remaining terms is detailed in the following.
A general definition of $E_{\rm{ext}}$, $E_{\rm{H}}$ and $E_{\rm{i-i}}$
can be found in App. \ref{app:DFT}. 
Those definitions stay valid provided one uses the cutoff potentials 
$\bar{V}_{\rm{ext}}$, $\bar{V}_{\rm{H}}$ and $\bar{\Phi}$. 
Unlike the potentials, the total energy is not defined up to a constant.
The choice of the $\bar{v}_c(\boG=0)$ can affect the value
of the energy contributions, but it should not affect the total energy.
In the following, choosing $\bar{v}_c(\boG=0)=0$ means that 
the $\boG=0$ terms of the {\it long range} contributions to the energy 
will be put to zero in the 2D framework. This is justified in App. \ref{app:2DQE}.

\subsubsection{External Energy}
The external energy $E_{\rm{ext}}$ is calculated via the sum of the 
eigenvalues of the KS system. 
Once the external potential is defined as in Eq. \ref{eq:extPot}, 
it is used to solve the KS system and the sum of its eigenvalues includes
the correct cutoff energy contribution.
The external energy decomposes into the energies of the electrons in the
potentials of: i) the ions $E^{\rm{ion}}_{\rm{ext}}$, ii) the gate 
$E^{\rm{gate}}_{\rm{ext}}$ and iii) the barrier $E^{\rm{barrier}}_{\rm{ext}}$.
We use sufficiently sharp and high barrier potentials to write that 
$E^{\rm{barrier}}_{\rm{ext}}\approx 0$
because there are (almost) no electrons where there is a potential barrier.
The other contributions $E^{\rm{ion}}_{\rm{ext}}$ and 
$E^{\rm{gate}}_{\rm{ext}}$ are non-zero, 
but we have no further modifications to make.

\subsubsection{Hartree Energy}
The Hartree energy is easily written in reciprocal space as:
\begin{align}
E_{\rm{H}} &=
\frac{\Omega}{2} \sum_{\mathbf{G} } n^*(\mathbf{G})  \bar{V}_{\rm{H}}(\mathbf{G}) ,  
\end{align} 
and is computed in practice by replacing $\bar{V}_{\rm{H}}(\boG)$ by its expression (Eq. \eqref{eq:barVh}):
\begin{align}
E_{\rm{H}} &=
\frac{\Omega}{2} \sum_{\mathbf{G} } |n(\mathbf{G})|^2  \bar{v}_c(\mathbf{G})  .
\end{align} 

\subsubsection{Ion-ion Interaction Energy}
The ion-ion interaction energy $E_{\rm{i-i}}$ is computed using the ion-ion interaction
potential $\bar{\Phi}$. The computation is based on the Ewald 
summation technique \cite{Ewald1921}, which involves a separation
into SR and LR parts $E_{\rm{i-i}}=E^{\rm{SR}}_{\rm{i-i}}+E^{\rm{LR}}_{\rm{i-i}}$.
Much like for the ionic potential, we do not need to modify the SR part. 
Here again, we start by presenting what is done in the original code, 
identify what we must modify, then present the implementation of the cutoff.

\underline{In the original 3D code}, following the Ewald summation technique,the ion-ion interaction potential $\Phi$ is separated in SR and LR part as follows:
\begin{align}  
\label{eq:Ewald}           
\Phi(\mathbf{r})
=& \Phi^{\rm{SR}}(\mathbf{r})+ \Phi^{\rm{LR}}(\mathbf{r}) - \Phi^{\rm{self}}  \\
\begin{split}
=&\sum_{\mathbf{R'}} \sum_{a'} {}^{'} \frac{ e^2 Z_{a'}}{|\mathbf{r}-\mathbf{R'}-\mathbf{d}_{a'}|}
\rm{erfc}(\sqrt{\eta_{\rm{ew}}}|\mathbf{r}-\mathbf{R'}-\mathbf{d}_{a'}|) \\
 &+
 \sum_{\mathbf{R'}} \sum_{a'}   \frac{e^2 Z_{a'}}{|\mathbf{r}-\mathbf{R'}-\mathbf{d}_{a'}|}
\rm{erf}(\sqrt{\eta_{\rm{ew}}}|\mathbf{r}-\mathbf{R'}-\mathbf{d}_{a'}|) \\
&- \Phi^{\rm{self}}  ,
\end{split}
\end{align}
where the prime in the first sum excludes the case $\{ \mathbf{R'=R}, a'=a \}$ if 
$\bor=\boR+\mathbf{d}_a$
and $\Phi^{\rm{self}}$ subtracts that term from the second sum.
The constant $\eta_{\rm{ew}}$ tunes the SR/LR separation 
(see App. \ref{app:eta} for more details).
The SR part of the ion-ion interaction
potential $\Phi^{\rm{SR}}$
is dealt with in real space and does not need to be modified as long as
${\rm{erf}}(\sqrt{\eta_{\rm{ew}}}l_z) \approx 1$ (easily satisfied).
$\Phi^{\rm{self}}$ is simply the value of $\Phi^{\rm{LR}}(\bor)$ for 
$\bor-\boR'-\mathbf{d}_{a'}=\bo{0}$. As such $\Phi^{\rm{self}}$ is also short-range.
We include the corresponding energy contributions in $E^{\rm{SR}}_{\rm{i-i}}$.
Those contributions do not need to be cut off.
The contribution of the LR potential $\Phi^{\rm{LR}}$ to the energy is 
computed in reciprocal space and needs to be be modified.

\underline{In our implementation}, we replace the Fourier transform of the 
LR part of the ion-ion interaction potential by its cutoff version: 
\begin{align}
\bar{\Phi}^{\rm{LR}}(\mathbf{G})&= 
\frac{1}{\Omega} \sum_a Z_a e^{i \mathbf{G} \cdot \mathbf{d}_a} 
\bar{v}_c(\mathbf{G}) e^{-|\mathbf{G}|^2/4\eta_{\rm{ew}}} .
\end{align} 
The LR contribution to the ion-ion interaction energy is then computed 
in reciprocal space as follows: 
\begin{align}
E_{\rm{i-i}}^{\rm{LR}}&=  \frac{\Omega}{2} \sum_{\mathbf{G}} n^*_{\rm{ion}}(\mathbf{G}) \bar{\Phi}^{\rm{LR}}(\mathbf{G}) \\
&=\frac{1}{2\Omega} \sum_{\boG} \left| \sum_a Z_a e^{i \mathbf{G} \cdot \mathbf{d}_a} \right|^2 
\bar{v}_c(\mathbf{G}) e^{-|\mathbf{G}|^2/4\eta_{\rm{ew}}} ,
\end{align} 
where $n_{\rm{ion}}(\boG)=\frac{1}{\Omega}\sum_a Z_a e^{i \boG \cdot \mathbf{d}_a}$
is the Fourier transform of the distribution of ions ($\varrho_{\rm{ion}}=en_{\rm{ion}}$).

\subsubsection{Other Energies}
The other energies to account for are the energy of the ions in the potential of the gates $E_{\rm{g-i}}$,
and the self interaction of the gates $E_{\rm{g-g}}$. 
\begin{align}
E_{\rm{g-i}} =&  \int_{\Omega} d\mathbf{r} \ n_{\rm{ion}}(\mathbf{r}) (-\bar{V}_{\rm{gate}}(z)) \\
 =& \sum_a Z_a 2\pi e^2 n_{\rm{bot}} \left(- |d_{a, z}-z^{\rm{bot}}_{\rm{g}}|+\frac{l_z}{2} \right) \\
 & +\sum_a Z_a 2\pi e^2 n_{\rm{top}} \left(- |d_{a, z}-z^{\rm{top}}_{\rm{g}}|+\frac{l_z}{2} \right) \\ 
E_{\rm{g-g}} =&  \frac{1}{2}\int_{\Omega} d\mathbf{r}  
\left( n_{\rm{bot}}\delta(z-z^{\rm{bot}}_{\rm{g}})+ n_{\rm{top}}\delta(z-z^{\rm{top}}_{\rm{g}})\right) \nonumber \\
& \ \ \ \times (-\bar{V}_{\rm{gate}}(z) ) \\
=&S \ \left(n_{\rm{bot}}+n_{\rm{top}}\right)^2 \pi e^2 \frac{l_z}{2}  \\
& +
S n_{\rm{bot}}n_{\rm{top}}  2 \pi e^2 
\left(- |z^{\rm{top}}_{\rm{g}}-z^{\rm{bot}}_{\rm{g}}| +\frac{l_z}{2} \right)
,
\end{align}
where $d_{a,z}$ is the $z$-component of $\mathbf{d}_a$.
We consider the most general case of a double-gate setup. 
Note that those contributions to the energy have a manifest dependency 
on the cutoff distance $l_z$. 
The total energy, of course, should not depend on $l_z$. 
As detailed in App. \ref{app:2DQE}, the $l_z$-dependent terms in the expression
above will cancel with corresponding terms in $E_{\rm{ext}}$ and $E_{\rm{H}}$.

\subsubsection{Verifications}
\begin{figure}[h]
\centering
\includegraphics[width=0.45\textwidth]{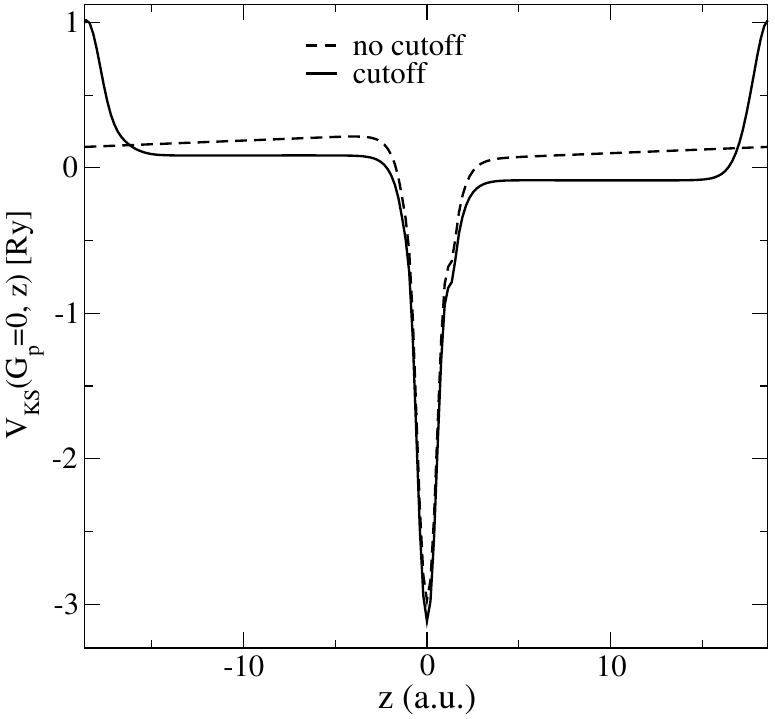}
\includegraphics[width=0.45\textwidth]{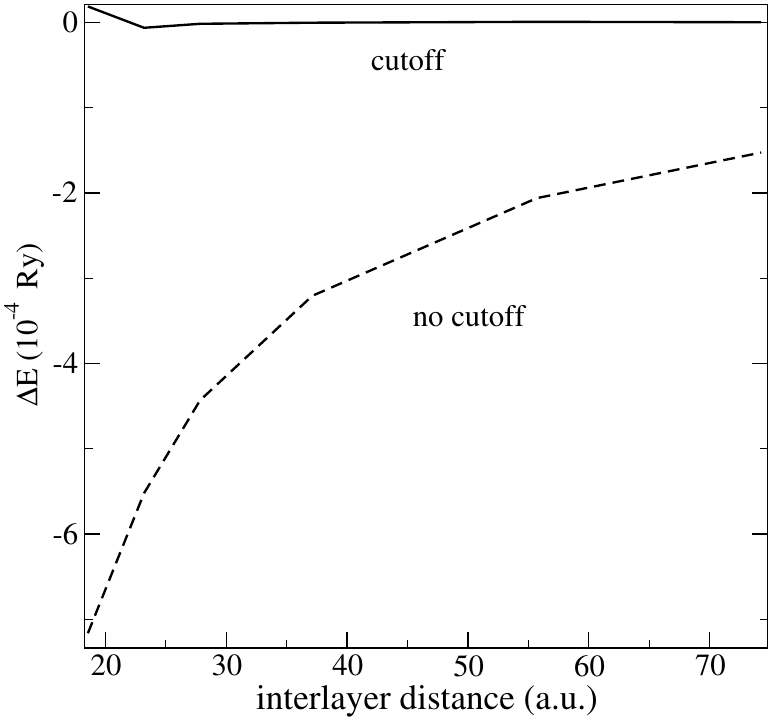}
\caption{DFT simulation of graphene with hydrogen atoms on top ($z>0$) of half the 
carbon atoms. The upper panel shows the planar-averaged KS potential, 
with and without using the 2D Coulomb cutoff.
The lower panel shows the total energy per unit cell as a function of the 
distance between periodic images, with and without using the 2D Coulomb cutoff. 
The zero for the energies corresponds to the total energy per unit cell obtained 
with cutoff.}
\label{fig:C-H}
\end{figure}

In the absence of doping, gate and barrier, the total energy is: 
\begin{align}
E^{\rm{neutral}}_{\rm{tot}} 
&= E_{\rm{kin}}+ E^{\rm{ion}}_{\rm{ext}}+E_{\rm{H}}+E_{\rm{XC}}+E_{\rm{i-i}}  .
\end{align}
We can thus check the consistency of the implementation of 
$E^{\rm{ion}}_{\rm{ext}}$, $E_{\rm{H}}$ and $E_{\rm{i-i}}$ in a neutral system.
We first compute the total energy of the neutral, non-polar system of Fig. \ref{fig:potentials}
with and without cutoff. We should obtain the same result as there is no issue with the periodic images
in that case. We checked that the difference is below numerical precision.
We can then use a neutral system with an out-of-plane dipolar moment 
such that interactions between periodic images
do play a role without the 2D Coulomb cutoff.
We use graphene with hydrogen atoms on top of half of the carbon atoms, see Fig. 
\ref{fig:C-H}. The effect of the Coulomb cutoff is clear on both the KS potential and total 
energy. 
The KS potential of the system without cutoff illustrates the comments
of Sec. \ref{sec:perio_im}. Namely, imposing 3D PBC leads to the compensation
of the out-of-plane dipolar moment by an external electric field, visible here
via the finite slope of the KS potential away from the material.
When we use the 2D Coulomb cutoff, we observe the right behavior, with a potential shift 
and no external electric field.
The energy of the system simulated without cutoff 
tends to the one with cutoff at large distances between periodic images.
With the cutoff, the energy is independent of the distance. 
There is a lower limit to the distance between periodic images, which is when the 
boundaries of the physical region are too close to the material. Still, the minimal
distance we can use in our implementation of the code is negligible with respect
to what we would have to use without cutoff. In the case of Fig. 
\ref{fig:C-H}, we see that the distance between the periodic images would have to be 
roughly five times larger without the cutoff to obtain the same total energy
as with the cutoff within $10^{-4}$ Ry. The computational cost would 
also be 5 times larger.
A way to get the right total energy in this kind of polar material is to simulate 
the mirror image of the system within the unit cell. We checked that this leads to
the same energy as what we find with the 2D Coulomb cutoff. Still, adding a mirror
image of the system rather than the cutoff leads to a drastic increase of the 
computational cost.

\subsection{Forces}

The forces on the ions are found by computing the derivative of the total energy with
respect to a displacement $\mathbf{u}_{a,i}$ of atom $a$ in direction $i$.
Only the terms related to an interaction in which the ions are involved remain.
The force acting on ion $a$ in direction $i$ is written:
\begin{align}
\label{eq:force}
\mathbf{F}_{a,i}= & -  \frac{\partial E_{\rm{tot}} }{\partial \mathbf{u}_{a,i}} 
=- \int_{\Omega} n(\mathbf{r}) \frac{\partial \bar{V}_{\rm{ion}}}{\partial \mathbf{u}_{a,i}} d\mathbf{r} 
-  \frac{\partial E_{\rm{i-i}}}{\partial \mathbf{u}_{a,i}} \\
& + \mathbf{F}^{\rm{g-i}}_{a,i} + \mathbf{F}^{\rm{b-i}}_{a,i} ,
\end{align}
where the first term is the force on the ion from the electrons, the second is from
the other ions, the third is from the gate and the last from the barrier. The notation 
$\frac{\partial }{\partial \mathbf{u}_{a,i}}$
implies taking the derivative at zero displacement $\mathbf{u}_{a,i}=0$.
All the quantities involved are known once the self-consistent calculation
is completed. We will only detail the terms for which we need to apply the 2D Coulomb cutoff.
The first term is calculated by computing the derivative of the ionic potential, separated 
in local and non-local parts.
The derivative of the local part is found by using the Fourier transform of the pseudopotentials:
\begin{align}
\frac{\partial \bar{V}^{\rm{loc}}_{\rm{ion}}(\mathbf{r})}{\partial \mathbf{u}_{a,i}} 
 &= -i \sum_{\mathbf{G}} \left( v_a^{ \rm{SR}}(\mathbf{G}) + 
\bar{v}_a^{ \rm{LR}}(\mathbf{G}) \right)
G_i  e^{- i \mathbf{G} \cdot \mathbf{d}_{a}}     
e^{i \mathbf{G} \cdot \mathbf{r}} .
\end{align}
The effect of the derivative in reciprocal space is to
bring down a factor $-iG_i$ from the exponential.  
The corresponding force is then calculated in reciprocal space:
\begin{align}
\begin{split}
-\int_{\Omega} n(\mathbf{r}) 
\frac{\partial \bar{V}^{\rm{loc}}_{\rm{ion}}(\bor)}{\partial \mathbf{u}_{a,i}} d\mathbf{r}
= & i\Omega \sum_{\mathbf{G} } n^*(\mathbf{G}) \times \\
& \left( v_a^{ \rm{SR}}(\mathbf{G}) + 
\bar{v}_a^{ \rm{LR}}(\mathbf{G}) \right)  
G_i  e^{- i \mathbf{G} \cdot \mathbf{d}_{a}} . 
\end{split}
\end{align}
The gate and the barrier have indirect contributions to this term. Indeed, 
they have an effect on $n(\mathbf{r})$, via their presence in the 
self-consistent KS potential.

The second term in Eq. \eqref{eq:force} is the force from the other ions. It is found
by derivation of the ion-ion interaction energy.
We only treat the LR contribution, because it is the only one that needs to be cut off:
\begin{align}
-  \frac{\partial E^{\rm{LR}}_{\rm{i-i}} }{\partial \mathbf{u}_{a,i}} =&   
-  \frac{\partial  }{\partial \mathbf{u}_{a,i}} \left(
\frac{\Omega}{2} \sum_{\mathbf{G}} n^*_{\rm{ion}}(\mathbf{G}) \bar{\Phi}^{\rm{LR}}(\mathbf{G})
\right) \\
=&
- \frac{1}{ \Omega} 
 \sum_{\mathbf{G}} \bar{v}_c(\mathbf{G}) e^{-|\mathbf{G}|^2/4\eta_{\rm{ew}}}  Z_a G_i  \\  
 & \times \sum_{a'} Z_{a'} \sin(\mathbf{d}_{a'}- \mathbf{d}_{a}) .
\end{align} 

The third term is the direct contribution of the electrostatic force 
applied by the gates to the ions.
Depending on doping, it can be repulsive or attractive:
\begin{align}
\mathbf{F}^{\rm{g-i}}_{a,z} = -  \frac{\partial E_{\rm{g-i}}  }{\partial \mathbf{u}_{a, z}} 
=& +Z_{a}  2\pi e^2 n_{\rm{bot}}  \  {\rm{sign}}(d_{a, z}-z^{\rm{bot}}_{\rm{g}}) \\
& +Z_{a}  2\pi e^2 n_{\rm{top}}  \  {\rm{sign}}(d_{a, z}-z^{\rm{top}}_{\rm{g}}) .
\end{align}
where we consider the most general case of the double-gate setup.

\begin{figure}[h]
\centering
\includegraphics[width=0.5\textwidth]{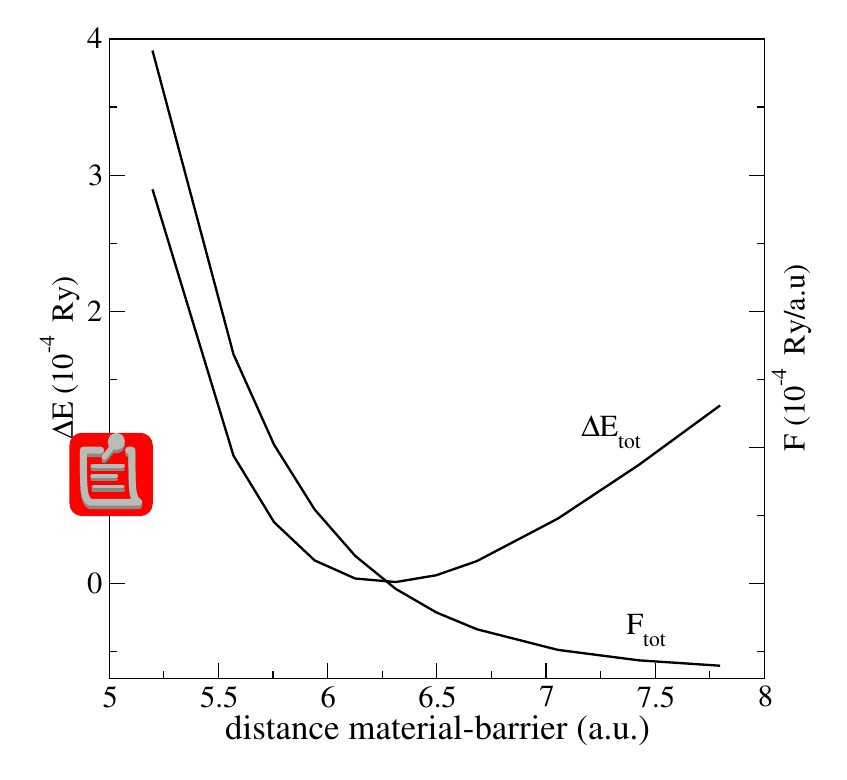}
\caption{ Total energy (variation from relaxed position) 
and forces as a function of the distance between the barrier and the 
2D material. The gate is moved with the barrier 
$z_{\rm{g}}=z_{\rm{b1}}-0.02c$ where $c$
is the distance between periodic images $c \approx 37$ a.u.
The position of the second barrier is such that it covers the unphysical region $z_{\rm{b2}}=z_{\rm{g}}+0.5c$.
The 2D material is graphene doped at a Fermi level of $+0.5$ eV.
The relaxed material-barrier distance is found to be around $6.26$ a.u.
 }
\label{fig:E-F_SB}
\end{figure}
The barrier applies no direct force on the ions $\mathbf{F}^{\rm{b-i}}_{a,i}=\mathbf{0}$. 
The ions effectively never see the barrier potential 
(they could if the barrier was smoother).
However, barriers can act on the ground-state
electronic density $n(\mathbf{r})$ which in turn acts on the ions. For example, 
if the system is too close from a barrier, the repulsive effect of the barrier 
will show in the self-consistent cycles, shifting the electrons away from the 
barrier.
The first term in Eq. \eqref{eq:force} will thus include a force that tends 
to push the ions away from the barrier, with the electrons.
The barrier is then essential to relax the forces, arriving at 
an equilibrium between the attraction from the gate and repulsion from the barrier. 
This is illustrated by the energy and forces of the gate-barrier-material system 
represented in Fig. \ref{fig:E-F_SB}. For large material-barrier distances, 
the force tends to the attraction from the gate, and the total energy is linear.
In that case, we have checked that the total force on the ions is the force between 
two charged plates, Eq. \eqref{eq:F_capa}. 
\begin{align}
\sum_a \mathbf{F}_{a,z}&=S \times 2\pi e^2 n^2_{dop} 
 \  {\rm{sign}}(z_{\rm{g}}) .
\end{align}
For small material-barrier distances, as the system gets too close to the barrier, 
there is a sharper increase of the force and the total energy. 

\subsection{Stresses for neutral 2D materials}

Stresses were not implemented in the FET setup.
We focus here on planar stress for neutral materials without external 
electric field, as it used in the relaxation of cell parameters for isolated 
2D materials.  
The following thus applies for Cartesian coordinates $i,j\in [x,y]$.
The stress is the first derivative of the energy with respect to strain\cite{Nielsen1985,Nielsen1985a}
\begin{align}
\mathbf{\sigma}_{i,j}= - \frac{1}{\Omega} \frac{\partial E_{tot}}{\partial \mathbf{\epsilon}_{i,j}}
\end{align}
For the full expressions of stresses, we refer the reader to 
Refs. \onlinecite{Nielsen1985,Nielsen1985a}, 
noting that the authors' definition of stress differs from QE's 
and ours by a minus sign. 
We only need to cutoff the terms involving long-range potentials. 
The expressions of those terms in the 3D framework are given in App. \ref{app:DFT}, 
along with a minimal outline of their derivation. They are modified as follows.

The key difference with the standard 3D code comes from the cutoff Coulomb 
interaction Eq. \ref{eq:CoulCut}, and its  differentiation with
respect to strain:
\begin{align}
\frac{\partial \bar{v}_c(\boG)}{\partial\epsilon_{ij}} &= 
\bar{v}_c(\boG)\frac{2G_iG_j}{\boG^2} (1-\beta(\boG_p,G_z))
\\
\begin{split}
\beta(\boG_p \neq 0,G_z)&=\frac{|\boG|^2 l_z}{2|\boG_p|} 
\frac{ e^{-|\mathbf{G}_p| l_z}  \cos(G_z l_z) }{ 1 - e^{-|\mathbf{G}_p| l_z}  
\cos(G_z l_z) } \\
\beta(\boG_p = 0,G_z)&=0
\end{split}
\end{align}
The Hartree contribution to the stress tensor reads
\begin{align}
\begin{split}
\sigma^{\rm{H}}_{i,j}=& -
\frac{1}{2} \sum_{\mathbf{G} } |n(\mathbf{G})|^2 \bar{v}_c(\mathbf{G})
 \times \\
& \left(
\frac{2G_iG_j}{\boG^2}  \left[  1
 - \beta(\boG_p,G_z)  \right]  - \delta_{ij} \right)
\end{split}.
\end{align}
The contribution from the long-range part of the local part of the ion-electron
interaction is
\begin{align}
\begin{split}
\sigma^{\rm{loc,LR}}_{i,j}=& -
\sum_{\mathbf{G}} n(\mathbf{G}) \sum_a e^{-i\boG \cdot \mathbf{d}_a } \bar{v}_a^{\rm{LR}}(\mathbf{G})
 \times \\
& \left(
\frac{2G_iG_j}{\boG^2}  \left[  1 + \frac{|\boG|^2}{4\eta}
 - \beta(\boG_p,G_z)  \right]  - \delta_{ij} \right)
\end{split}.
\end{align}
Finally, the long-range contribution from the ion-ion interaction gives:
\begin{align}
\begin{split}
\sigma^{\rm{i-i,LR}}_{i,j}=& - \frac{1}{2 \Omega}
\sum_{\mathbf{G}} \left| \sum_a Z_a e^{i\boG \cdot \mathbf{d}_a } \right|^2 \bar{v}_c(\mathbf{G}) e^{|\boG|^2/4\eta_{\rm{ew}}}
 \times \\
& \left(
\frac{2G_iG_j}{\boG^2}  \left[  1 + \frac{|\boG|^2}{4\eta_{\rm{ew}}}
 - \beta(\boG_p,G_z)  \right]  - \delta_{ij} \right)
\end{split}
\end{align}

\subsection{Phonons and EPC}

To calculate the phonon dispersion and electron-phonon interactions, 
as reminded in App. \ref{app:DFPT}, 
we need to compute the response of the electronic density to a phonon 
perturbation.
In essence, the linear response of the system involves derivatives 
of the previous potentials and energies.
Once the previous framework is set up, 
we just have to apply the Coulomb cutoff to the derivatives consistently. 
The following applies to insulators, semiconductors and metals. In the latter case, 
Fermi-surface effects, and notably the shift of the Fermi level that may arise with phonons 
at zero momentum, are treated as in the standard code, see Ref. \onlinecite{Baroni}.

\subsubsection{Dynamical matrix}

In practice, the dynamical matrix is given by the following integrals
on the unit cell:
\begin{align} \label{eq:DynMat}
\begin{split}
\mathcal{D}_{a,i,a',j} \times & \sqrt{M_a M_{a'}} =  \int_{\Omega} d\mathbf{r} 
\frac{\partial^2 \bar{V}_{\rm{ion}}(\mathbf{r})}
{\partial \mathbf{u}_{a,i}(\mathbf{q})\partial \mathbf{u}_{a',j}(\mathbf{q})} n(\mathbf{r}) \\ 
&+
\int_{\Omega} d\mathbf{r} \left( \frac{\partial \mathcal{\bar{V}}_{\rm{ion}}(\mathbf{r})}
{\partial \mathbf{u}_{a,i}(\mathbf{q})}\right)^* 
\left( \frac{\partial n(\mathbf{r})}{\partial \mathbf{u}_{a',j}(\mathbf{q})} e^{-i\boq\cdot \bor}\right)\\
&+
\mathcal{D}^{\rm{i-i}}_{a,i,a',j}  
\end{split}
\end{align}
where the "$\mathcal{V}$" notation indicates that we are using the lattice periodic 
part of the potential, see App. \ref{app:DFPT}.
The first term can readily be computed from the quantities obtained in the 
ground-state calculation.
It is computed in reciprocal space as:
\begin{widetext}
\begin{align}
\begin{split}
\int_{\Omega} d\mathbf{r} &
\frac{\partial^2 \bar{V}^{\rm{loc}}_{\rm{ion}}(\mathbf{r})}
{\partial \mathbf{u}_{a,i}(\mathbf{q})\partial \mathbf{u}_{a',j}(\mathbf{q})} n(\mathbf{r}) = 
-\delta_{a,a'} \Omega \sum_{\boG} \left( v_a^{ \rm{SR}}(\mathbf{q}+\mathbf{G}) + 
\bar{v}_a^{ \rm{LR}}(\mathbf{q}+\mathbf{G}) \right) G_i G_j \Re \left( n^*(\boG) e^{-i\boG \cdot \mathbf{d}_a} \right) .
\end{split}
\end{align}
where $\Re(x)$ gives the real part of $x$.
The last term comes from the second derivative of the ion-ion interaction $E_{\rm{i-i}}$
with respect to a phonon displacement. The contribution from $E^{\rm{SR}}_{\rm{i-i}}$
does not change. The LR part
$E^{\rm{LR}}_{\rm{i-i}}$ yields the following contribution 
to the dynamical matrix:
\begin{align}
\begin{split}
\mathcal{D}^{\rm{i-i,LR}}_{a,i,a',j} =&
\frac{1}{\Omega} \sum_{\mathbf{G}, \mathbf{q}+\mathbf{G} \ne 0} 
\bar{v}_c(\mathbf{q}+\mathbf{G}) e^{- |\mathbf{q}+\mathbf{G}|^2/4\eta_{\rm{ew}}} 
Z_a Z_{a'} (\mathbf{q}+\mathbf{G})_i (\mathbf{q}+\mathbf{G})_j 
e^{i (\mathbf{q}+\mathbf{G}) \cdot (\mathbf{d}_a-\mathbf{d}_{a'})} \\
& -
\frac{1}{\Omega} \sum_{\mathbf{G} \ne 0} \bar{v}_c(\mathbf{G}) e^{- |\mathbf{G}|^2/4\eta_{\rm{ew}}} 
Z_a  G_i G_j  \left( \sum_{a''} Z_{a''} 
\cos(\mathbf{G}\cdot(\mathbf{d}_a-\mathbf{d}_{a''})) \right) \delta_{a, a'} .
\end{split}
\end{align}
\end{widetext}
The second term in Eq. \eqref{eq:DynMat} is computed 
via numerical integration over the unit cell in real space.
The quantities inside the integral are computed during the 
calculation of the electronic density response to the perturbed KS potential, 
presented in the following.

\subsubsection{Perturbed KS potential}

The linear electronic density response is found by solving a self-consistent system
involving the effective perturbation, that is the derivative of the KS potential
with respect to a phonon displacement
$\frac{\partial \mathcal{\bar{V}}_{\rm{KS}}(\mathbf{r}_p, z)}{\partial \mathbf{u}_{a,i}(\mathbf{q})}$
, Eq. \eqref{eq:pertKS} (the notation "$\mathcal{\bar{V}}$" indicates 
lattice periodic functions, see App. \ref{app:DFPT}).
The first term is the perturbation of the external potential. 
The phonons only bring a direct perturbation to the potentials in 
which the ions are involved. 
This means the perturbed external potential contains only the contribution from
the ionic potential. 
The Fourier transform of the derivative of the local part of the ionic potential 
has non-zero components 
at wave vectors $\boq+\boG$:
\begin{align} \label{eq:dViondu}
\frac{\partial \mathcal{\bar{V}}^{\rm{loc}}_{\rm{ion}}(\mathbf{q}+\mathbf{G})}{\partial \mathbf{u}_{a,i}(\mathbf{q})}
=& -i \left( v_a^{ \rm{SR}}(\mathbf{q}+\mathbf{G}) + 
\bar{v}_a^{ \rm{LR}}(\mathbf{q}+\mathbf{G}) \right) \\
& \times (\mathbf{q}+\mathbf{G})_i 
e^{-i(\mathbf{q}+\mathbf{G})\cdot \mathbf{d}_a} , 
\end{align}
where the Fourier components of the long-range part of the local pseudopotential are 
similar to Eq. \eqref{eq:pseudoLRG}:
\begin{align}
\bar{v}_a^{ \rm{LR}}(\mathbf{q}+\mathbf{G}) = 
- \frac{Z_s}{\Omega} \bar{v}_c(\mathbf{q}+\mathbf{G}) 
e^{-|\mathbf{q}+\mathbf{G}|^2/4\eta} .
\end{align}
The perturbed ionic potential of Eq. \eqref{eq:dViondu} 
(along with the non-local part that is computed as in the original 3D code),
is Fourier transformed and inserted in the second term of the dynamical matrix Eq. \eqref{eq:DynMat}.

The remaining long-range potential to cut off is the Hartree potential
generated by the density response, computed in reciprocal space: 
\begin{align}
\frac{\partial \mathcal{\bar{V}}_{\rm{H}}(\mathbf{q}+\mathbf{G})}{\partial \mathbf{u}_{a,i}(\mathbf{q})} &=
\bar{v}_c(\mathbf{q}+\mathbf{G}) 
\frac{\partial n(\mathbf{q}+\mathbf{G})}{\partial \mathbf{u}_{a,i}(\mathbf{q})} .
\end{align}

The density response, solution of the self-consistent system corresponding to the effective perturbation
$\frac{\partial \mathcal{V}_{\rm{KS}}(\mathbf{r}_p, z)}{\partial \mathbf{u}_{a,i}(\mathbf{q})}$
is inserted in the second term of the dynamical matrix Eq. \eqref{eq:DynMat}.

\subsubsection{Born Effective charges and LO-TO splitting}

In polar materials, the long-wavelength behavior of longitudinal optical (LO) modes 
depends strongly on dimensionality. Indeed, the displacement patterns of LO phonons are 
asociated with dipoles that interact with each others via long-range 
Coulomb interactions. These dipole-dipole interactions lead to an extra term in the energy of the LO mode with respect to the tranverse optical (TO) mode, thus leading to the so-called LO-TO splitting.
In 2D, as shown in Ref. \onlinecite{Sohier2017a}, 
the splitting vanishes in the zero momentum limit, but the dispersion of the LO mode 
displays a finite slope at the $\mathbf{\Gamma}$ point.
The implementation of the 2D cutoff in DFPT as detailed 
above guarantees the correct treatment of the LO-TO splitting.
A key quantity for this phenomenon is the tensor of Born effective charges. 
Notably, it gives the values of the finite slope of the LO dispersion at zero momentum. 
It can by computed either via the forces induced by an electric field perturbation, or 
via the polarization induced by atomic displacements. 
In both cases, the quantities involved (forces, perturbed KS potential) 
are already corrected as detailed above.

\subsubsection{Fourier interpolation of phonon dispersions}

Dynamical matrices can be Fourier interpolated\cite{Gonze1997,Baroni} 
to obtain phonons on dense grids at minimal computational cost. 
The Fourier interpolation in itself is carried out as in the standard 3D 
code.
In polar materials, however, the LO-TO splitting corresponds to 
a discontinuity either in the zeroth (in 3D) or first (in 2D) order derivative of the 
phonon dispersion. Some non-analytic terms arise at long wavelengths due to the long-
range nature of the dipole-dipole interactions. Those non-analytic terms must be modeled 
and excluded from the interpolation process. Since they depend on dimensionality, 
the interpolation requires a different treatment in 3D and 2D. The 2D treatment is 
implemented as detailed in Ref \onlinecite{Sohier2017a}.

\subsubsection{EPC}
We have all the quantities necessary to compute the EPC:
\begin{align}\label{eq:EPC}
g_{\mathbf{k}+\mathbf{q},s,\mathbf{k},s',\nu}
 &= \sum_{a,i} \mathbf{e}^{a,i}_{\boq,\nu} \sqrt{\frac{\hbar}{2M_a \omega_{\mathbf{q},\nu}}}  \langle  \mathbf{k+q},s |
 \frac{\partial \mathcal{\bar{V}}_{\rm{KS}}(\mathbf{r})}{\partial \mathbf{u}_{a,i}(\mathbf{q}) } |\mathbf{k},s' \rangle .
\end{align}
The EPC matrix elements are screened via the induced part of the effective KS 
perturbation (Hartree and exchange-correlation). 
The Hartree part of the screening is then that of a 2D material.
The gate and the barrier have no direct effect in the KS perturbation. Note that
they are absent from this section. However, they broke the symmetry of the 
ground-state. In particular, the electronic distribution is not centered on the 
ions' plane anymore. We will study the consequences of their presence in the 
following section.

\section{Application to Graphene FET setup}

In this section we exploit our implementation
of DFPT for gated 2D systems to simulate
some predicted peculiarities of the FET setup. 
For isolated graphene without any external electric field, 
it can be shown that the flexural ZA phonons disperse quadratically 
and their energy is zero in the long-wavelength limit\cite{Katsnelson2012}. 
Based on the mirror symmetry with respect to the graphene plane, 
one can further show that ZA phonons do not couple 
linearly to electrons\cite{Manes2007}. 
Those characteristics do not hold for graphene in the FET setup.
First, the phonon dispersion changes due to the presence of a substrate 
and a gate-dielectric. Second, the presence of an electric field breaks the mirror 
symmetry with respect to the graphene plane, making linear coupling to electrons 
possible. Those FET-specific effects have not been studied in 
the context of DFPT.
The electron-phonon coupling with flexural phonons in gated graphene
was recently studied by first-principles 
and suggested to be a significant scattering mechanism\cite{Gunst2017}. 
However, in this work, the calculations performed do not
completely include the effect of metallic screening on the
electron-phonon coupling. Indeed, at the two lowest doping considered,
the phonon momentum allowed for by $11\times 11$ supercell is too large
with respect to the size of the Fermi surface. 
Furthermore, the method used in Ref. \onlinecite{Gunst2017}
assumes the Fourier transform of the
derivative of the self-consistent potential to be phonon-momentum independent.
We will show that this is not the case in doped graphene.

We perform DFPT calculations on graphene doped in the FET setup.
We simulate the main consequences of the presence of the substrate 
and gate dielectric by placing two barriers at 
$z_{b1}$ and $z_{b2}=-z_{b1}$, such that $z_{b2}-z_{b1}\approx 5.3$ \AA. 
Compared to graphite, 
it corresponds to a graphene-barrier 
distance that is smaller than the distance between two adjacent 
graphene atomic planes, but larger than the distance separating the tails of the 
electronic densities associated with those planes.
In practice, the distances between graphene and the substrate 
or the dielectric depend on the details of the interactions between 
those materials. Here, we simply make a choice.
As we will see, the results of the phonon calculations point to a rather 
conservative choice for the graphene-barrier distance.
We plot the KS potential of the system in the upper panel of Fig. 
\ref{fig:FET-phonons}. We will consider three setups:
\begin{itemize}
\item{``constant field"}: we use two oppositely charged gates. There is a finite and constant electric field between the gates, but graphene stays neutral, i.e. $\varepsilon_F=0$ eV with respect to the Dirac point. 
In this setup, the mirror symmetry associated to the graphene plane is broken
and electronic screening from graphene is minimized. Barriers are present.
\item{``one gate"}: we use a single bottom gate ($z_{\rm{bot}}<0$), 
with a charge equal and opposite to that of the graphene sheet. 
The electric field is finite on the side of the bottom gate, but 
zero on the side of the top gate. 
In this setup the mirror symmetry associated to the graphene plane is also broken 
but the graphene is electron-doped such that $\varepsilon_F \approx 0.7$ eV. 
This implies a stronger metallic screening from $\pi^*$ electrons. Barriers are present.
\item{``isolated"}: this setup is simply for comparison. We simulate isolated, neutral graphene, without gates and without barriers. 
\end{itemize}
Linear response calculations are performed within the Quantum ESPRESSO 
distribution\cite{Giannozzi2009}, using the 2D cutoff and FET setup 
implementation described in this work. 
We use a norm conserving pseudopotential within 
the local density approximation \cite{Perdew1981} (LDA).
A dense k-point grid ($96\times 96 \times 1$) is chosen 
to sample the Fermi surface of graphene and account for screening effects.
We use a $0.01$ Ry Methfessel-Paxton smearing function for the electronic 
integration and a $65$ Ry kinetic energy cutoff. 
We use the relaxed equilibrium structure of isolated graphene in all setups. 
We neglect the change in lattice parameter due to doping, which was calculated to be 
under $0.1\%$. In the out-of-plane direction, the graphene sheet is fixed midway between 
the barriers, where the repulsive forces from the barriers cancel out. We neglect the 
comparatively small attractive forces from the charged gates.

\subsection{Acoustic out-of-plane (ZA) phonons}

\begin{figure}[h!]
\centering
\includegraphics[width=0.45\textwidth]{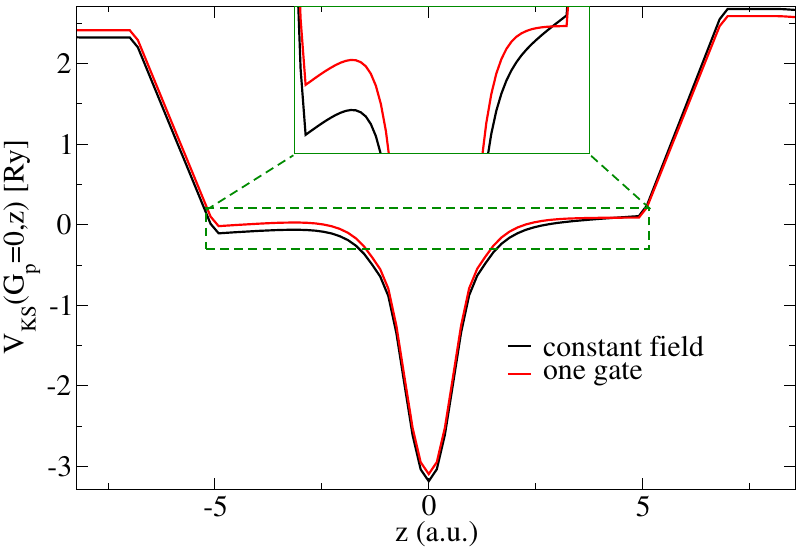}
\includegraphics[width=0.47\textwidth]{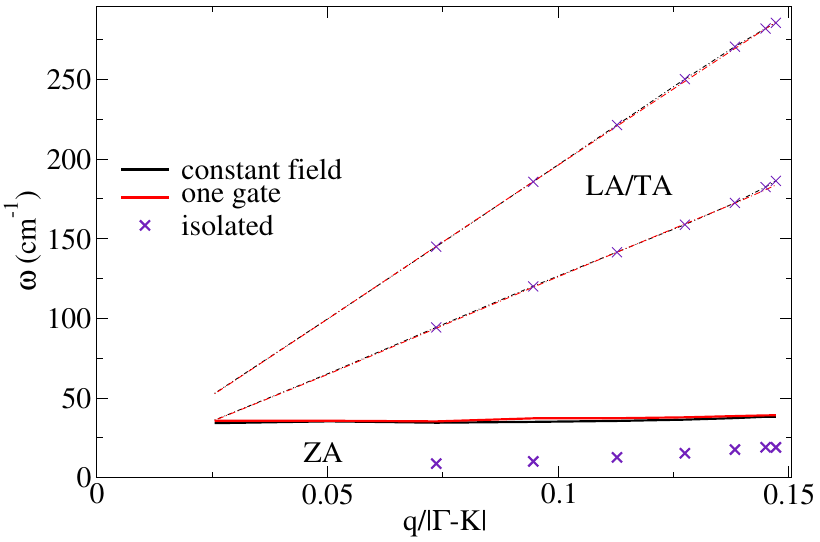}
\caption{On the top is the KS potential of the simulated system. 
The inset zooms on the dashed rectangle to highlight the difference between 
the setups setups. In the ``constant field" setup, the slope of the KS potential, and thus the electric field, is the same on both sides of the graphene layer.
Graphene is neutral. 
In the ``one gate" setup, 
the electric field vanishes on one side and graphene is electron-doped.
Below is the dispersion of the acoustic phonons for the different setups. We are mainly interested in the dispersion of ZA phonons, but the in-plane acoustic modes are shown in thin dash-dot lines for comparison.}
\label{fig:FET-phonons}
\end{figure}
We first show the emergence of a finite 
ZA phonon frequency at $\mathbf{\Gamma}$ when graphene is enclosed 
between two barriers. We calculate the acoustic phonons in this system 
and get the dispersion in the lower panel of Fig. \ref{fig:FET-phonons}. 
The dispersions obtained for isolated graphene (no gate, no barrier) 
are also shown. 
The dispersion of the in-plane modes is essentially unaffected by the 
presence of gates and barriers.
In contrast, a shift in the dispersion of the ZA phonons is observed 
for the ``one gate" and ``constant field" setups. The shift being similar for both setups, it can be attributed to the presence of barriers rather than the electric field configuration.
We observe in this case a rather flat dispersion, with 
$\omega_{\rm{ZA}}(\mathbf{\Gamma})\approx 35 $ cm$^{-1}$.
When a 2D material is enclosed between two potential barriers, the ZA phonon dispersion loses its quadratic behavior in the long wavelength limit. 
Instead, it goes to a finite value at $\mathbf{\Gamma}$. The closer the barriers, 
the more confined is the 2D material and the larger is 
$\omega_{\rm{ZA}}(\mathbf{\Gamma})$.
For graphene, a relevant reference for that value might be the ZO' mode 
of graphite, in which neighboring layers are in out-of-phase ZA modes.
It is often found\cite{Nicklow1972,Oshima1988,Mounet2005} to have a 
$\mathbf{\Gamma}$ frequency close to $100$ cm$^{-1}$.
The relatively small value of $\omega_{\rm{ZA}}(\mathbf{\Gamma})$ found here 
would thus indicate that the chosen graphene-barrier distance is rather 
conservative in the sense that the effect is most likely underestimated. 
Such a situation is preferred here, in order to find an upper bound for 
the strength of scattering by ZA phonons.

\subsection{Gate-induced coupling to ZA phonons}
We now demonstrate the emergence of a finite coupling 
to linear order between the electrons and out-of-plane 
acoustic ZA phonons, due to the electric field breaking the mirror symmetry 
with respect to the graphene plane. More importantly, we unravel the 
critical impact of screening on this coupling.
We consider scattering of electrons on an iso-energetic line at 
$\varepsilon=0.7$ eV in the $\pi^*$ band. 
In the ``one gate" setup, this corresponds to the Fermi surface of 
graphene. Thus, the results will be representative of the scattering involved in 
electronic transport. We use the same iso-energetic line in the ``constant field" 
setup, although the line does not represent the Fermi surface since the 
graphene layer is neutral. In this situation, the results are not relevant for 
electronic transport. They correspond to the relaxation of a single 
electron excited at an energy of $\varepsilon=0.7$ eV. 
The motivation behind comparing scattering on the same iso-energetic lines
is to observe the effect of electronic screening.
\begin{figure}[h]
\centering
\includegraphics[width=0.47\textwidth]{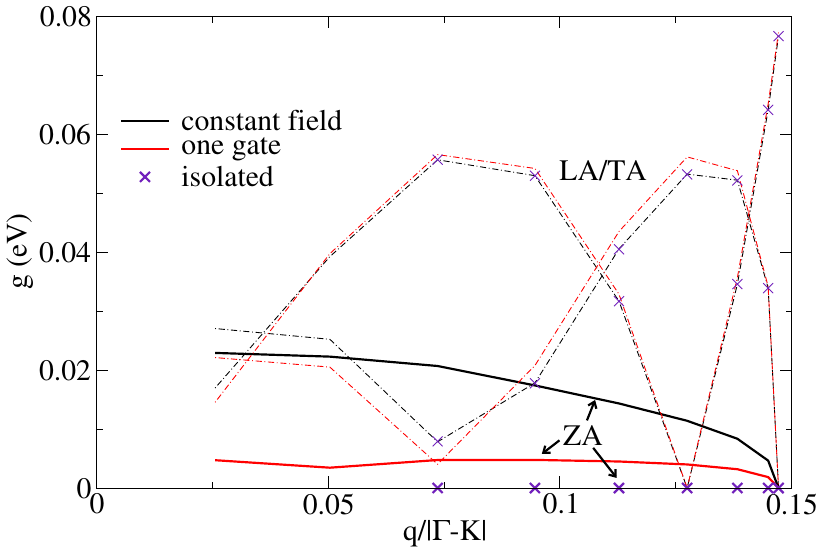}
\includegraphics[width=0.47\textwidth]{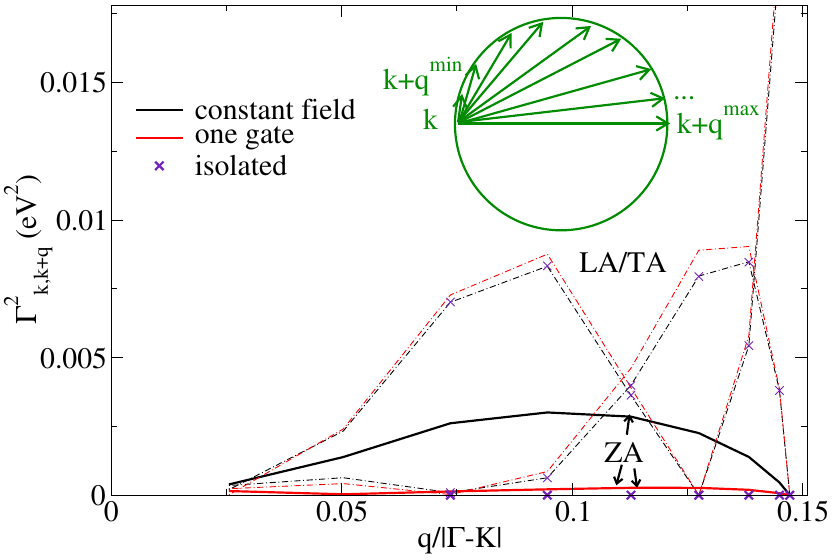}
\caption{ Emergence of a finite coupling to ZA phonons for graphene under an 
electric field. The upper panel shows EPC matrix elements while 
the lower panel shows 
the transport-relevant quantity defined in Eq. \ref{eq:Gkkq}. 
The set of pairs of electronic states involved in the scattering processes are
chosen such that the scattered states span half the Fermi surface,    
as represented in green. In-plane phonons are represented in thin 
dash-dot lines for comparison. The ``constant field" setup, in which graphene is 
neutral, leads to ZA couplings that are comparable to the in-plane phonons. 
The ``one gate" setup, in which graphene is doped, leads to negligible 
ZA couplings. 
}
\label{fig:gZA}
\end{figure}
We fix the initial state $|\bok\rangle$ and define a set of scattered states 
$|\bok+\boq\rangle$ on the iso-energetic line that we assume circular, 
as represented in Fig. \ref{fig:gZA}. 
This implies that we neglect trigonal warping and assume elastic 
scattering, that is 
$\varepsilon_{\bok+\boq}=\varepsilon_{\bok} \pm \hbar \omega_{\boq,\nu}
\approx \varepsilon_{\bok}$. 
We cover only half the line, the other half being equivalent by symmetry. 
We plot the corresponding EPC as a function of the norm of momenta. 
Keep in mind that increasing norm then corresponds to increasing values of
the angle between initial and scattered states $\theta_{\bok+\boq,\bok}$.

We plot the corresponding values of 
$g_{\nu}(\boq)=g_{\bok,\pi^*, \bok+\boq,\pi^*,\nu}$ for $\nu=$ZA, TA, LA 
in the upper panel of Fig. \ref{fig:gZA}. The in-plane acoustic modes are plotted 
for comparison. We clearly observe the emergence of a non-zero value for 
$g_{\rm{ZA}}$, as expected. In both neutral and doped cases, 
the value of the coupling vanishes for the large phonon momenta 
corresponding to backscattering.
In the neutral case, $g_{\rm{ZA}}$ is comparable to the coupling with 
in-plane acoustic phonons. 
However, while the coupling to in-plane phonons is essentially doping-independent, 
$g_{\rm{ZA}}$ is much smaller in doped graphene. Indeed, the coupling to in-plane phonon
is dominated by  gauge fields\cite{Sohier2014a}, which do not affect the local charge 
density and thus are not screened. In contrast, 
the gate-induced coupling to ZA phonon acts as a deformation potential, 
that is a periodic modulation of charge density in the underlying 
lattice potential in which the free electrons move. 
In terms of Hamiltonian, the perturbation is diagonal and proportional to  
identity in the Dirac spinor basis. As such, this perturbation 
is screened by graphene's electrons. In the doped case, 
the gate-induced coupling $g_{\rm{ZA}}$ is strongly screened 
by metallic graphene and it becomes negligible.
In Eq. (2) of Ref. \onlinecite{Gunst2017}, the authors considered a 
deformation potential (called "field-induced coupling constant") that depends on 
gate-voltage but is independent of momentum $\boq$. The screening from 
the conduction electrons of graphene does in fact bring a strong momentum 
dependency\cite{Sohier2015} to this quantity.

\begin{figure}[h]
\centering
\includegraphics[width=0.47\textwidth]{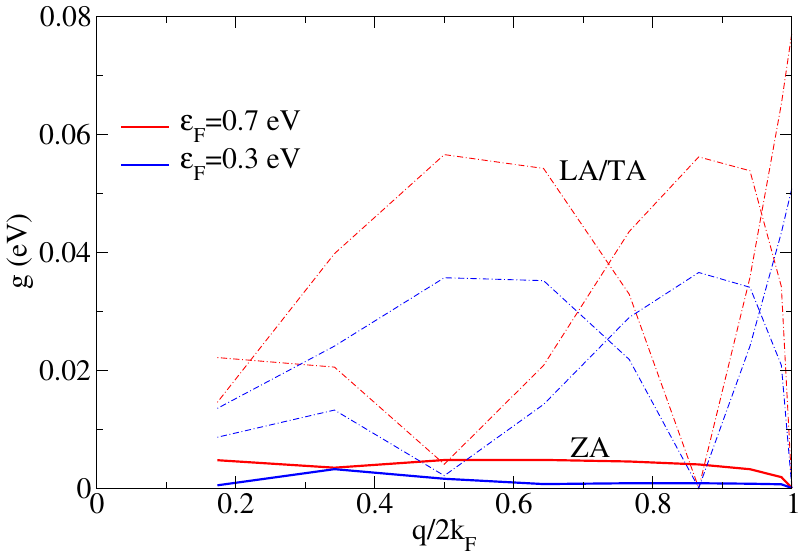}
\caption{ The field-induced coupling with flexural phonons decreases when doping decreases. The electronic pairs involved in the scattered processes are chosen as in Fig. \ref{fig:gZA}, with an iso-energetic line taken at the corresponding Fermi levels $\varepsilon_F=0.7$ eV and $\varepsilon_F=0.3$ eV, respectively.
}
\label{fig:dopdep}
\end{figure}

In Ref. \onlinecite{Gunst2017}, it is argued that despite relatively 
weak coupling, the high occupation of ZA phonons lead to a considerable 
scattering probability. In the lower panel of Fig. \ref{fig:gZA}, 
we study the transport-relevant quantity 
\begin{align}\label{eq:Gkkq}
\mathbf{\Gamma}^2_{\bok, \bok+\boq,\nu}=g^2_{\nu}(\boq) (1+2N_{\boq, \nu}) (1-\cos(\theta_{\bok+\boq,\bok})) 
\end{align} 
where $N_{\boq, \nu}$ is the phonon occupation at room temperature 
(B\"ose-Einstein distribution with $T=300$ K) and the angular term conveys the fact 
that backscattering is more detrimental to electronic transport. 
In the framework of the relaxation time approximation and elastic processes, the integral of this term over the Fermi surface corresponds to the scattering rate.
In the neutral case, we see in the lower panel of Fig. \ref{fig:gZA}
that despite vanishing backscattering and the 
relatively small coupling overall, $\mathbf{\Gamma}_{\bok, \bok+\boq,\rm{ZA}}$ is 
comparable to the other acoustic mode, 
thanks to a relatively large occupation of ZA 
phonons (note that as mentioned before, the phonon frequency is probably a lower 
bound so the occupation and the scattering rate are upper bounds).  This makes 
field-induced scattering by ZA phonons potentially important for carrier 
relaxation in neutral graphene under a constant electric field.
In the more relevant case of single-gated and doped graphene, 
the scattering from the ZA phonon is screened and negligible.
The doping level considered here is rather large.
However, for lower, more experimentally realistic doping levels, 
the coupling would be similar or smaller.
Indeed, the electric field and thus the field-induced 
bare coupling would be smaller. Since the screening scales with the dimension of 
the Fermi surface \cite{Sohier2015}, similar screening would be obtained for 
scattering around the Fermi surface at any doping. This general trend is verified 
in Fig. \ref{fig:dopdep}, where we use two different doping and compare the 
electron-phonon coupling as a function of the momentum rescaled by 
the size of the Fermi surface $|\boq|/2k_F$.
Finally, in addition to screening effects, the flatness of the 
ZA dispersion also plays a role in decreasing the coupling with respect to the quadratic dispersion of isolated graphene (see definition Eq. \ref{eq:EPC}). In experimental setups, the ZA dispersion will depend on the details of the interactions with the subtrate and the gate dielectric. In our simulations, the barriers act as an approximation to these interactions. However, as mentionned above, we expect
experimental values of $\omega_{\rm{ZA}}(\mathbf{\Gamma})$ to be comparable or larger than the simulated value.
Thus, we expect the values of the coupling found in the highly 
doped case to be an upper bound for scattering around the Fermi 
surface of graphene at any finite doping level. This coupling 
would thus be undetectable in transport measurements.
Those results are in stark contrast with the recent first-principles
study of this effect\cite{Gunst2017}. 
The main reasons for this discrepency are the dispersion of flexural phonons 
and electronic screening. Due to the flat dispersion of the flexural phonons, we find 
overall smaller couplings to flexural phonons than in Ref. \onlinecite{Gunst2017}. For 
the relaxation of photoexcited carriers in neutral graphene ("constant field" setup), we 
find that the coupling is still comparable to the coupling with in-plane phonons. 
For carrier transport in doped graphene, however, the coupling to flexural phonons is 
screened and becomes negligible.
Electronic screening plays a key role in 
electron scattering. It is a highly dimensionality-dependent quantity 
that requires the correct 2D framework. It can be quite difficult
to model and predict its effect on complex mechanisms like 
electron-phonon coupling, even more so in complex systems like 
gated 2D heterostructures. As illustrated here, adequate DFPT methods 
are then an invaluable tool.

\section{Conclusion}
Manipulating the electronic properties of heterostructures via the field effect is 
key to many future usage of 2D materials. We first set the framework for the 
simulation of charged heterostructures within the field-effect setup. We then show 
that various issues arise within the standard three-dimensional periodic boundary 
conditions for those systems, and that the truncation of the Coulomb interaction in 
the out-of-plane direction is a simple and efficient solution. We detail the 
implementation of the two-dimensional Coulomb cutoff and the field effect 
setup within the Quantum ESPRESSO distribution, for ground state and
linear response calculations. 
The most basic changes concern the construction of potentials 
equivalent to the those generated by an isolated two-dimensional system. 
Changes are then made accordingly throughout the code to compute physical 
quantities properly defined in the two-dimensional framework. 
This includes total energies, forces, stresses, phonons and 
electron-phonon interactions. We demonstrate the relevance of the 
implementation by studying flexural (or out-of-plane acoustic) phonons 
for graphene in the field effect setup. 
Our results show the emergence of a finite phonon frequency in the long 
wavelength limit, as well as a finite coupling to electrons. 
However, electronic screening makes the coupling to flexural phonons
negligible with respect to the coupling to in-plane phonons.
This implies that the phenomenon is undetectable in transport measurements 
for graphene at finite doping.

\section{acknowledgements}
The authors acknowledge support from the European Union Horizon 2020 research and innovation program under Grant agreement No. 696656-GrapheneCore1 and from
Agence Nationale de la Recherche under the reference n. ANR-13-IS10-0003-01. 
Part of the calculations were carried out at IDRIS, CINES, CEA and BSC TGCC.

\appendix
\section{DFT}
\label{app:DFT}
In this section we introduce the quantities that can be calculated in DFT.
We give the most straightforward expressions for a generic system in 3D space.
The formulas might be applied to  3D-periodic or 2D-periodic materials. 
The following description is obviously very far from exhaustive. 
The aim is simply to set the notations and provide definitions for the quantities 
mentioned in the main text.
For more details, we refer the reader to the literature, for example Refs. 
\onlinecite{Martin2004,Engel2011,Parr2015}.
The lattice vectors are noted $\boR$, the internal coordinate of atom $a$ is $\mathbf{d}_a$.
The reciprocal lattice vectors $\boG$ are such that $e^{i \boG \cdot \boR}=1$.

\subsection{Potentials}

The central potential in the self-consistent process is the 
Kohn-Sham\cite{Kohn1965} potential.
It contains:
\begin{itemize}
\item $V_{\rm{ext}}$: the principal contribution to the external potential 
$V_{\rm{ext}}$ is the potential generated by the ions 
$V_{\rm{ion}}$ calculated via the pseudopotentials. For the purpose of this paper, 
we consider only the local part of $V_{\rm{ion}}$, written as:
\begin{align}
V^{\rm{loc}}_{\rm{ion}}(\bor)=\sum_{\boR,a} v_a(\bor-\boR-\mathbf{d}_a)
\end{align}
where $v_a$ is the pseudopotential associated to atom $a$.
We can put other contributions
into $V_{\rm{ext}}$, but in this appendix $V_{\rm{ext}}=V_{\rm{ion}}$.

\item $V_{\rm{H}}$:
the Hartree potential is given by:
\begin{align}
V_{\rm{H}}(\mathbf{r}) = e^2 \int d\mathbf{r'} 
\frac{n(\mathbf{r'})}{|\mathbf{r}-\mathbf{r'}|},
\end{align}
Here, and whenever there is no specified interval, the integrals are carried out over 
the entire space spanned by the corresponding variable.
\item $V_{\rm{XC}}$: the exchange-correlation potential 
is based on the local density approximation \cite{Perdew1981} (LDA). 
\end{itemize}

We also have to mention the potential $\Phi(\bor)$ which is the potential generated 
by the ions used to compute ion-ion interactions. 
Although it is generated by the same source as $V_{\rm{ion}}$, 
it is not calculated via the pseudopotentials. To generate this potential, 
the ions are modeled by a collection of point charges. 
This potential is defined as:
\begin{align}
\Phi(\bor)={\sum_{\mathbf{R'}, a'}}^{'} 
\frac{ e^2 Z_{a'}}{|\bor-\mathbf{R'}-\mathbf{d'}_{a}|}
\end{align}
where the prime on the sum excludes the $\{\boR=\boR', a=a' \}$ case
if $\bor=\boR+\mathbf{d}_a$.

\subsection{Total Energy}

What we call the total energy of the system is the clamped-ions energy or the Born-Oppenheimer energy surface \cite{Born1927}:
\begin{align}
\mathcal{E}_{\rm{tot}} = \mathcal{E}_{\rm{kin}}+ \mathcal{E}_{\rm{XC}}+
\mathcal{E}_{\rm{H}} + \mathcal{E}_{\rm{ext}}+ \mathcal{E}_{\rm{i-i}}
\label{eq:E_tot}
\end{align}
where
\begin{align}
\mathcal{E}_{\rm{kin}}&=
 -  \frac{\hbar^2}{ 2m_e} \sum_{\bok,s} f(\varepsilon_{\bok,s}) \langle \psi_{\bok,s} | \nabla^2  | \psi_{\bok,s} \rangle \\
\mathcal{E}_{\rm{H}} &=
 \frac{1}{2} \int  n(\mathbf{r})V_{\rm{H}}(\bor) d\mathbf{r}
 \\
\mathcal{E}_{\rm{ext}}&=
 \int n(\mathbf{r})V_{\rm{ext}}(\bor) d\mathbf{r} \\
\mathcal{E}_{\rm{i-i}} &= 
  \frac{1}{2} \sum_{\mathbf{R}, a}  Z_{a} \Phi(\mathbf{R}+\mathbf{d}_{a})
\end{align}
and $\mathcal{E}_{\rm{XC}}$ is the exchange-correlation energy.
For any of the above quantity $\mathcal{E}$, one can define the corresponding energy per unit cell
$E=\mathcal{E}/N$ where $N$ is the number of unit cells. This is often more useful in practice, 
since the system is infinite, as are the energies $\mathcal{E}$.

\subsection{Forces}
To calculate the force on atom $a$ in direction $i$, 
we compute the derivative of the total energy per unit cell
with respect to a displacement $\mathbf{u}_{a,i}$ of this atom in this direction, 
and take the value at $\mathbf{u}_{a,i}=0$.
Using the Hellmann-Feynman theorem \cite{Feynman1939,Hellmann1937}, 
the force acting on ion $a$, in direction $i$ is given by:
\begin{align}
\label{eq:forces2Dmat}
\mathbf{F}_{a,i}&=-  \frac{\partial E_{\rm{tot}}}{\partial \mathbf{u}_{a,i}} \\
&=- \int_{\Omega}     
n(\mathbf{r}) \frac{\partial V_{\rm{ext}} (\mathbf{r}) }
{\partial \mathbf{u}_{a,i}}   d\bor
-  \frac{\partial E_{\rm{i-i}}}{\partial \mathbf{u}_{a,i}} ,
\end{align} 
where the integral is carried over the volume of the unit cell $\Omega$.
Here and in the following, the notation $\frac{\partial }{\partial \mathbf{u}_{a,i}}$ 
represents the value of the derivative at zero displacement.
The first term is the contribution from the electrons, the second from the ions. 
The forces can be computed as soon as we have solved the ground state, 
since there are only known quantities and their derivatives.

\subsection{Stresses}
The stress is the first derivative of the energy with respect 
to the strain tensor $\overleftrightarrow{\epsilon}$
\cite{Nielsen1985,Nielsen1985a}:
\begin{align}
\mathbf{\sigma}_{i,j}= - \frac{1}{\Omega} \frac{\partial E_{\rm{tot}}}{\partial \mathbf{\epsilon}_{i,j}}
\end{align}
In practice, it is derived from the total energy by applying the scaling procedure 
$\bor' = (1+\overleftrightarrow{\epsilon}) \bor$.
We will focus here on the contributions from long-range potentials, the only ones involved in the 2D Coulomb cutoff process.
We give an outline of the general derivation, below is the treatment 
of the  Hartree contribution:
\begin{align}
\mathbf{\sigma_{i,j}^{\rm{H}}} &= - \frac{1}{\Omega} \frac{\partial E_{\rm{H}}}{\partial \epsilon_{ij}} \\
\text{with } \  E_{H}&=\frac{\Omega}{2} \sum_{\boG \ne 0} n^*(\boG) V_{\rm{H}}(\boG)
\end{align}
Since $\Omega n(\boG)$ is the number of electrons per unit cell, 
it is invariant under strain. We are thus left with the derivative of the Hartree potential:
\begin{align}
\frac{\partial V_{\rm{H}}(\boG)}{\partial \epsilon_{ij}} &= 
\frac{\partial v_c (\boG)}{\partial \epsilon_{ij}} n(\boG)  
+
\frac{\partial n(\boG)}{\partial \epsilon_{ij}} v_c (\boG)
\end{align}
Starting from the fact that $\Omega n(\boG)$ is invariant and 
knowing that the volume transforms as $\Omega' \approx (1+\sum_i \epsilon_{ii}) \Omega$, one finds that $\frac{\partial n(\boG)}{\partial \epsilon_{ij}}=-\delta_{ij}n(\boG)$. 
We are left with the derivative of the Coulomb interaction.
Since the scaling procedure gives 
$\boG'= (1-\overleftrightarrow{\epsilon})\boG$ in reciprocal space,
we have that $\frac{\partial G_l}{\partial \epsilon_{ij}}=-\delta_{li} G_j$.
The derivative of the Coulomb interaction is then computed using the chain rule:
\begin{align}
\frac{\partial v_c(\boG)}{\partial\epsilon_{ij}}& = \sum_l 
\frac{\partial v_c(\boG)}{\partial G_l} \frac{\partial G_l}{\partial \epsilon_{ij}} 
\\
&= -\frac{\partial v_c(\boG)}{\partial G_i} G_j 
\\
&= v_c(\boG) \frac{2G_i G_j}{\boG^2}
\end{align}
with the 3D Coulomb interaction as $v_c(\boG)=\frac{4 \pi e^2}{\boG^2}$.
The Hartree contribution is thus:
\begin{align}
\begin{split}
\sigma^{\rm{H}}_{i,j}=& -
\frac{1}{2} \sum_{\mathbf{G} } |n(\mathbf{G})|^2 v_c(\mathbf{G})
 \times \left(
\frac{2G_iG_j}{\boG^2}  - \delta_{ij} \right)
\end{split}.
\end{align}
Similar derivations yield the contribution from the long-range part of 
the local ionic potential. By differentiation of (see also Eq. \ref{eq:LR-SR3D}):
\begin{align}
E_{\rm{ext}}^{\rm{loc, LR}}&=\Omega\sum_{\boG} n^*(\boG)\sum_a e^{i\boG \cdot  \mathbf{d}_a} 
v^{\rm{LR}}_{a}(\boG)
\end{align}
one gets:
\begin{align}
\begin{split}
\sigma^{\rm{loc,LR}}_{i,j}=& -
\sum_{\mathbf{G}} n(\mathbf{G}) \sum_a e^{-i\boG \cdot \mathbf{d}_a } 
v_a^{\rm{LR}}(\mathbf{G})
 \times \\
& \left(
\frac{2G_iG_j}{\boG^2}  \left[  1 + \frac{|\boG|^2}{4\eta} \right]  
- \delta_{ij} \right)
\end{split}.
\end{align}
Finally, the long-range part of the ion-ion interaction is written:
\begin{align}
E_{\rm{i-i}}^{\rm{LR}}&= 
\frac{1}{2\Omega} \sum_{\boG} \left| \sum_a Z_a e^{i \mathbf{G} \cdot \mathbf{d}_a} \right|^2 
v_c(\mathbf{G}) e^{-|\mathbf{G}|^2/4\eta_{\rm{ew}}} ,
\end{align} 
and gives the following stress:
\begin{align}
\begin{split}
\sigma^{\rm{i-i,LR}}_{i,j}=& - \frac{1}{2 \Omega}
\sum_{\mathbf{G}} \left| \sum_a Z_a e^{i\boG \cdot \mathbf{d}_a } \right|^2 v_c(\mathbf{G}) e^{|\boG|^2/4\eta_{\rm{ew}}}
 \times \\
& \left(
\frac{2G_iG_j}{\boG^2}  \left[  1 + \frac{|\boG|^2}{4\eta_{\rm{ew}}} \right]  
- \delta_{ij} \right)
\end{split}
\end{align}

\section{DFPT}
\label{app:DFPT}
DFPT enables the computation of the linear response of the ground 
state to given perturbation. We focus here on phonon perturbations, 
as implemented in the phonon code of the Quantum ESPRESSO distribution.
Again, we only give minimal description to settle 
the notation and define the quantities referred to in the main text. 
The reader might refer to the literature for more details, for example Refs. 
\onlinecite{Gonze1997,Baroni}.

\subsection{Phonons}
A phonon perturbation of momentum $\mathbf{q}$ is represented by a collection of
displacements $\mathbf{u}_{a,i}(\mathbf{R})$ of atom $a$ in Cartesian direction
$i$: 
\begin{align}
\mathbf{u}_{a,i}(\mathbf{R}) = \mathbf{u}_{a, i}
(\mathbf{q}) e^{i \mathbf{q} \cdot \mathbf{R} } 
\end{align}
where $\mathbf{u}_{a, i}(\mathbf{q})$ is the Fourier transform of  $\mathbf{u}_{a, i}(\mathbf{R})$.
The phonon frequencies are obtained from the second derivative of the total energy of the 
crystal $\mathcal{E}_{\rm{tot}}$ (not the energy of a unit cell) via the
matrix of the interatomic force constants defined as \cite{Giannozzi1991,Gonze1997,Baroni}:
\begin{align}
\mathbf{C}_{ai, a'j} (\mathbf{R}-\mathbf{R}') &= \frac{\partial^2 \mathcal{E}_{\rm{tot}}}
{\partial \mathbf{u}_{a,i}(\mathbf{R}) \partial \mathbf{u}_{a',j}(\mathbf{R}')} \\
&= \mathbf{C}^{ion}_{ai, a'j} (\mathbf{R}-\mathbf{R}') +
\mathbf{C}^{elec}_{ai, a'j} (\mathbf{R}-\mathbf{R}') \nonumber
\end{align}
In this particular context, 
it does not make sense to talk about energy per unit cell. Indeed, the energy 
is not lattice periodic because of the phonon perturbation.
There are two contributions, one from the electrons, 
one from the ions:
\begin{align}
\label{eq:Celec}
\mathbf{C}^{elec}_{ai, a'j} (\mathbf{R}-\mathbf{R}')=&
\int  \frac{\partial^2 V_{\rm{ext}}(\mathbf{r})}
{\partial \mathbf{u}_{a,i}(\mathbf{R})\partial \mathbf{u}_{a',j}(\mathbf{R'}) } n(\mathbf{r}) d\mathbf{r} \\
& + \int  \frac{\partial V_{\rm{ext}}(\mathbf{r})}{\partial \mathbf{u}_{a,i}(\mathbf{R})} 
\frac{\partial n(\mathbf{r})}{\partial \mathbf{u}_{a',j}(\mathbf{R'})}d\mathbf{r} \\
\label{eq:Cions}
\mathbf{C}^{ions}_{ai, a'j} (\mathbf{R}-\mathbf{R}') &=
\frac{\partial^2 \mathcal{E}_{i-i}}{\partial \mathbf{u}_{a,i}(\mathbf{R}) \partial  \mathbf{u}_{a',j}(\mathbf{R'})}
\end{align}
The first term of Eq. \eqref{eq:Celec} and Eq. \eqref{eq:Cions} 
are simply the second derivatives of quantities already computed in DFT. 
The second term of Eq. \eqref{eq:Celec}, however, contains 
the linear response of the electronic density  
to a phonon perturbation.
This quantity can be calculated within DFPT. 
A phonon perturbation translates into a periodic perturbation of the potential 
generated by the ions, that is a periodic perturbation of $V_{\rm{ext}}$:
\begin{align} 
\frac{\partial V_{\rm{ext}}(\mathbf{r})}{\partial \mathbf{u}_{a, i} (\mathbf{q}) }=\frac{\partial V_{\rm{ion}}(\mathbf{r})}{\partial \mathbf{u}_{a, i} (\mathbf{q}) }
\end{align}
where we now work with the (single-component) Fourier transform of the phonon perturbation 
$\mathbf{u}_{a, i} (\mathbf{q})$.
The phonon perturbation triggers the linear response of the electronic density:
\begin{align}
\frac{\partial n(\mathbf{r})}{\partial \mathbf{u}_{a, i} (\mathbf{q}) }
\label{eq:dndu},
\end{align}
which is found by solving a new set of equations, 
involving the linear perturbation to the KS potential:
\begin{align} \label{eq:pertKS}
\frac{\partial \mathcal{V}_{\rm{KS}}(\mathbf{r})}{\partial \mathbf{u}_{a,i}(\mathbf{q})}= 
\frac{\partial \mathcal{V}_{\rm{ext}}(\mathbf{r})}{\partial \mathbf{u}_{a,i}(\mathbf{q})} 
+
\frac{\partial \mathcal{V}_{\rm{H}}(\mathbf{r})}{\partial \mathbf{u}_{a,i}(\mathbf{q})} +
\frac{\partial \mathcal{V}_{\rm{XC}}(\mathbf{r})}{\partial \mathbf{u}_{a,i}(\mathbf{q})} 
\end{align}
where the $\mathcal{V}$ notation indicates that we take the lattice-periodic part 
of the perturbations:
\begin{align}
\frac{\partial V(\mathbf{r})}{\partial \mathbf{u}_{a,i}(\mathbf{q})} =
\frac{\partial \mathcal{V}(\mathbf{r})}{\partial \mathbf{u}_{a,i}(\mathbf{q})} e^{i \boq \cdot \bor}.
\end{align}
The last two terms of Eq. \eqref{eq:pertKS} are generated by 
the density response Eq. \eqref{eq:dndu}. 
We thus have a new self-consistent system to solve.
Once self-consistency is reached, we can calculate the dynamical matrix $\mathcal{D}$
which is the Fourier transform of the matrix
of the force constants:
\begin{align}
\mathcal{D}_{a,i,a',j} (\mathbf{q})
&= \frac{1}{ \sqrt{M_a M_{a'}} } \sum_{ \mathbf{R}}
\mathbf{C}_{ai, a'j} (\mathbf{R}) 
e^{i\mathbf{q}\cdot\mathbf{R}}
\end{align}
where $M_a$ is the mass of atom $a$, and we have used translational invariance to express
the matrix of the force constant as a function of the generic lattice vector $\mathbf{R}$.
The eigenvalue problem:
\begin{align}
\omega^2(\mathbf{q}) \mathbf{u}_{a,i}(\mathbf{q}) &=  \sum_{ a', j }
\mathcal{D}_{a,i,a',j} (\mathbf{q}) 
\mathbf{u}_{a,i}(\mathbf{q})
\end{align}
gives the frequencies $\omega_{\boq, \nu}$ ($\omega^2_{\boq, \nu}$ being the eigenvalues) and eigenvectors 
$\mathbf{e}_{\mathbf{q},\nu}$ of mode $\nu$ at momentum $\boq$.

\subsection{EPC}
The electron-phonon interaction matrix elements are obtained from the 
derivative of the KS potential as follows:
\begin{align}\label{eq:g2QE}
g_{\mathbf{k}+\mathbf{q},s,\mathbf{k},s',\nu}
 &= \sum_{a,i} \mathbf{e}^{a,i}_{\boq,\nu} \sqrt{\frac{\hbar}{2M_a \omega_{\mathbf{q},\nu}}}  \langle  \mathbf{k+q},s |
 \frac{\partial \mathcal{V}_{\rm{KS}}(\mathbf{r})}{\partial \mathbf{u}_{a,i}(\mathbf{q}) } |\mathbf{k},s' \rangle  
\end{align}

\section{Long-range / short range separations}
\label{app:eta}

There are two short-range / long-range (SR/LR) separations performed in the code. 
One for the pseudopotentials (Eq. \ref{eq:LR-SR3D}) and
one for the computation of the ion-ion interaction, Eq. \ref{eq:Ewald}. 
They are done for different reasons. The first is done to 
enable the computation of the Fourier transform of the pseudopotentials, 
the second is done to optimize the convergence of the real/reciprocal space 
computation of the Ewald sums. To each is associated a tuning parameter, 
$\eta$ or $\eta_{\rm{ew}}$.
Since in both case, the long-range/short-range contributions are 
always put back together before computing any physical quantity, 
the tuning parameters can be chosen separately. 
As in the original code, we set $\eta=1$ for the pseudopotentials while
the Ewald splitting parameter $\eta_{\rm{ew}}$ used in the computation of the ion-ion interaction is chosen depending on the plane-wave cutoff. 

\section{Treatment of the $\boG=0$ singularities in the 2D code}
\label{app:2DQE}
\subsection{$\boG=0$ value of the Coulomb interaction}

The treatment of the $\boG=0$ terms with a 2D Coulomb cutoff
is developed in Ref. \onlinecite{Rozzi2006}, where the authors 
show that one should use the following value for the $\boG \to 0$ limit
of the truncated Coulomb interaction:
\begin{align}
\bar{v}_{c}(\mathbf{G} \to 0)=- 2 \pi e^2 l_z^2 
\end{align}
In the following, we quickly justify this recommendation.
Note that we use the notation $\mathbf{G} \to 0$ to distinguish this value from 
the value $\bar{v}_{c}(\mathbf{G} = 0)=0$ used in our implementation. 
The potential $V(\bor_p, z)$ generated by 
a generic 2D distribution $m(\bor_p, z)$ via the cutoff Coulomb interaction is written: 
\begin{align}
V(\bor_p, z)= e^2 \int_{\rm{plane}} \int^{+l_z}_{-l_z} 
\frac{m(\bor'_p,z')}{\sqrt{|\bor'_p-\bor_p|^2+(z'-z)^2} } d\bor'_p dz' 
\end{align}
By changing variables and exploiting the in-plane periodicity
of $m(\bor_p, z)$, 
it can be shown that the planar average of the potential $V(\boG_p=0, z)$ 
reads:
\begin{align}
V(\boG_p=0, z) &= e^2 \int_{\rm{plane}} \int^{+l_z}_{-l_z} 
\frac{\langle m\rangle_p (z')}{\sqrt{|\bor_p|^2+(z'-z)^2} } 
d\bor_p dz' 
\end{align}
This can be written as :
\begin{align}
V(\boG_p=0, z) 
&= \int^{+l_z}_{-l_z}  \langle m\rangle_p (z')
\ \ \bar{v}_c(\boG_p=0, |z-z'|) \   
 dz' 
\end{align}
with 
\begin{align}
\bar{v}_c(\boG_p=0, |z|) &= e^2 \int_{\rm{plane}}  \frac{1}{\sqrt{|\bor_p|^2+z^2} }  d\bor_p  \\
&= e^2 \int_{\rm{plane}} \left[ \frac{1}{|\bor_p| } + \frac{1}{\sqrt{|\bor_p|^2+z^2} } -\frac{1}{|\bor_p| }\right] d\bor_p  \\
&=
e^2 \int_{\rm{plane}}  \frac{d\bor_p}{|\bor_p| } - 2 \pi e^2  |z|
\end{align}
The first term of the above equation is the one that gives the diverging behavior
in the potential of a charged plane.
However, this term vanishes as soon as the 2D system is 
globally neutral within the cutoff
because it does not depend on $z$. If we replace $m$ by a globally neutral 
distribution $n_{\rm{tot}}$ that would be 
the sum of the distributions of the electrons, ions and gate, we get:
\begin{align}
 \int^{+l_z}_{-l_z} & \langle m\rangle_p (z') \left( e^2 \int_{\rm{plane}}  \frac{d\bor_p}{|\bor_p| } \right) dz' = \\
&  \left( e^2 \int_{\rm{plane}}  \frac{d\bor_p}{|\bor_p| } \right) \int^{+l_z}_{-l_z}  \langle n_{\rm{tot}}\rangle_p (z')  dz' =0
\end{align}
We can thus drop this term.
The definition of the $\bar{v}_{c}(\mathbf{G} \to 0)$ is then found by Fourier 
transform of the remaining term along the third direction (as in Eq. \eqref{eq:2DTF}):
\begin{align}
\bar{v}_{c}(\mathbf{G} \to 0) &= \frac{1}{c} \int^{+l_z}_{-l_z}  \left( -2 \pi  e^2 |z| \right) dz = - 2 \pi e^2 l_z^2 
\end{align}

\subsection{Implementation}
We now show why we can further simplify the process and use $\bar{v}_{c}(\boG=0) =0$ in our implementation.
The exchange-correlation and barrier contributions to the potentials 
and energies are ignored here because they bring no divergence. 
Since we are considering cutoff quantities, 
the following concerns the long-range part of the potentials 
and the corresponding contributions to energy when the long-range/short-range separation is done.
In order to simplify the argument, we do not make the distinction in the notation. 
In the following, the "tilde" quantities are those defined using the value $\bar{v}_{c}(\mathbf{G} \to 0)=-2e^2\pi l_z^2$
recommended in Ref. \cite{Rozzi2006}. 
Here is how we define the potentials in our implementation:
\begin{align}
\begin{split}
\bar{V}_{\rm{H}}(\mathbf{r}) &= \tilde{V}_{\rm{H}}(\mathbf{r})- \bar{v}_c(\mathbf{G}\to 0)  n(\boG=0) \\
\bar{V}_{\rm{ion}}(\mathbf{r}) &= \tilde{V}_{\rm{ion}}(\mathbf{r})+ \bar{v}_c(\boG \to 0)  n_{\rm{ion}}(\boG=0) \\
\bar{V}_{\rm{gate}}(\mathbf{r}) &= \tilde{V}_{\rm{gate}}(\mathbf{r})+ 
\bar{v}_c(\boG \to 0)  \frac{n_{\rm{dop}}}{c} \\
\bar{\Phi}(\mathbf{r}) &= \tilde{\Phi}(\mathbf{r})- 
\bar{v}_c(\boG \to 0)  n_{\rm{ion}}(\boG=0) \\
\end{split}
\end{align}
Defined this way, the $\boG=0$ value of our potentials is zero 
(at least for the long-range part in the 
case of $\bar{V}_{\rm{ion}}$ and $\bar{\Phi}$). 
Note that if we sum $\bar{V}_{\rm{H}}$, $\bar{V}_{\rm{ion}}$ and $\bar{V}_{\rm{gate}}$, 
we find that $\bar{V}_{\rm{KS}}=\tilde{V}_{\rm{KS}}$, which is essential.
The potentials give the following energies:
\begin{align}
\begin{split}
E_{\rm{H}}  &= \tilde{E}_{\rm{H}} -
\frac{\Omega}{2}  (n_{\rm{ion}}(0)+\frac{n_{\rm{dop}}}{c})^2  \bar{v}_c(\boG \to 0) 
\\
E^{\rm{ion}}_{\rm{ext}}  &= \tilde{E}^{\rm{ion}}_{\rm{ext}} +
\Omega  (n_{\rm{ion}}(0)+\frac{n_{\rm{dop}}}{c})n_{\rm{ion}}(\boG=0)  \bar{v}_c(\boG \to 0)
  \\
E^{\rm{gate}}_{\rm{ext}}  &= \tilde{E}^{\rm{gate}}_{\rm{ext}} + 
\Omega  (n_{\rm{ion}}(\boG=0)+\frac{n_{\rm{dop}}}{c})\frac{n_{\rm{dop}}}{c}  \bar{v}_c(\boG \to 0)
\\
E_{\rm{i-i}}  &= \tilde{E}_{\rm{i-i}} -
\frac{\Omega}{2} n^2_{\rm{ion}}(\boG=0)  \bar{v}_c(\boG \to 0) 
\\
E_{\rm{g-i}}  &= \tilde{E}_{\rm{g-i}} -
\Omega \frac{n_{\rm{dop}}}{c}n_{\rm{ion}}(\boG=0)  \bar{v}_c(\boG \to 0)
\\
E_{\rm{g-g}}  &= \tilde{E}_{\rm{g-g}} - 
\frac{\Omega}{2} \left( \frac{n_{\rm{dop}}}{c} \right)^2  \bar{v}_c(\boG \to 0)
\end{split}
\end{align} 
where, once again, we have that all the $\boG=0$ contributions to the energy are 
zero, and that $E_{\rm{tot}}=\tilde{E}_{\rm{tot}}$, if we sum all the contributions.
The process described above is equivalent to setting $\bar{v}_c(\boG = 0)=0$, 
and it gives the same KS potential and total energy 
as using $\bar{v}_c(\boG \to 0)=-2 \pi e^2 l_z^2$. It is also very close to the 
process used in the original 3D code, which allows us to minimize the changes.

The $\boG=0$ values of the {\it cutoff} potentials 
can thus be set to zero. This is pretty straightforward to apply when the whole potential is cutoff, without prior SR/LR separation, that is in the case of 
$\bar{V}_{\rm{H}}$ and $\bar{V}_{\rm{gate}}$.
When the SR/LR separation is done, for $\bar{V}_{\rm{ion}}$ and 
$\bar{\Phi}$, it becomes more subtle, because the short-range parts
are not cutoff and the corresponding $\boG=0$ term are finite.
Let's examine what must be done, always trying to minimize the changes 
with respect to the 3D code.
The ionic potential's $\boG=0$ term in the 3D code is
\begin{align}
V_{\rm{ion}}(\boG=0) &= \tilde{V}_{\rm{ion}}(\boG=0)+ v_c(\boG=0)  n_{\rm{ion}}(\boG=0).
\end{align}
This term is non-divergent, non-zero and short-range. It is computed by
numerical integration in a finite sphere. It can be referred to as the 
"$\alpha$" term, leading to the so-called "$\alpha$Z" energy contribution. 
It is a combination of two divergent terms. 
Following the overall strategy of the 2D cutoff implementation, the first term  
should be separated in SR/LR part and then the LR part replaced by its cut off 
counterpart. The second term would be directly replaced by its cut off counterpart. 
Let us follow this process and determine how the "$\alpha$" term 
should be corrected in 2D. 
We first separate the 3D ionic potential in SR and LR parts. 
The SR part depends on the pseudopotential and it is left unchanged.
The LR part of the 3D potential $\tilde{V}^{3D, LR}_{\rm{ion}}(\boG=0)$ 
can be separated in two terms: 
(i) a divergent part and (ii) a finite, nonsingular part\cite{Martyna1999} 
\begin{align}
\tilde{V}^{3D,LR, NS}_{\rm{ion}}(0)&=- \frac{e^2}{\Omega}\sum_a Z_a
\int \frac{\rm{erf}(\sqrt{\eta}|\bor|)-1}{|\bor|} d\bor \\
&= \frac{\pi e^2}{\eta\Omega}\sum_a Z_a
\end{align}
The divergent part (i) cancels out with $v_c(\boG \to 0)  n_{\rm{ion}}(\boG=0)$ in 3D. It's cut off counterpart similarly cancels with $\bar{v}_c(\boG \to 0)  n_{\rm{ion}}(\boG=0)$ in 2D. The changes to be made to the 
"$\alpha$" term would thus be subtracting and adding zero, i.e, doing nothing.
The term (ii) in its 3D form, $\tilde{V}^{3D, LR, NS}_{\rm{ion}}(0)$ should be 
subtracted and replaced by the $\boG=0$ value of its cut off counterpart.
The latter being zero, the only correction to make to the 3D "$\alpha$" term is to subtract $\frac{\pi e^2}{\eta\Omega}\sum_a Z_a$.
In the 3D Ewald summation, the $\boG=0$ term is:
\begin{align}
\Phi(\boG=0) &= \tilde{\Phi}(\boG=0)- v_c(\boG \to 0)  n_{\rm{ion}}(\boG=0)
\end{align}
Similarly to the "$\alpha$" term, the only thing to do is to 
subtract $-\frac{\pi e^2}{\eta_{\rm{ew}}\Omega}\sum_a Z_a$.
Note that it looks like we are adding some $\eta$-dependent terms to $E^{\rm{ion}}_{\rm{ext}}$ and $E_{\rm{i-i}}$. This is actually not the case. 
The $\eta$-dependent terms we add in the $\boG=0$ terms cancel out with 
equal and opposite contributions in the $\boG \ne 0$ terms. We are
defining the $\boG = 0$ term that corresponds to the cutoff potential 
used for the $\boG \ne 0$ terms.
$E^{\rm{ion}}_{\rm{ext}}$ and $E_{\rm{i-i}}$ are independent of $\eta$ or 
$\eta_{\rm{ew}}$, and those parameters can be still be chosen independently 
for the two SR/LR separation processes as said in App. \ref{app:eta}.

\bibliographystyle{apsrev4-1}
\bibliography{2Dqe}

\end{document}